\documentclass[a4paper,twoside,12pt]{article}
\usepackage[pdftex]{graphicx}
\usepackage[T1]{fontenc}
\usepackage{lmodern} 
\usepackage[utf8]{inputenc}
\usepackage{multirow}
\usepackage{dcolumn}
\newcolumntype{d}[1]{D{.}{\cdot}{#1} }
\usepackage{indentfirst}
\usepackage{amsmath}
\usepackage{mathastext} 
\usepackage{t1enc} 
\usepackage{multirow}
\usepackage{float}
\usepackage{wrapfig}
\usepackage{enumerate}
\usepackage{caption}
\usepackage{subcaption}
\usepackage{color}
\usepackage{array}
\usepackage{chngpage}
\usepackage[top=4cm ,bottom=4cm ,left=3cm ,right=3cm]{geometry}
\usepackage{xcolor}
\usepackage{upgreek}
\DeclareGraphicsExtensions{.pdf,.eps,.png,.jpg,.mps,.gif}
\usepackage[unicode]{hyperref}
\hypersetup{
  colorlinks,
  citecolor=black,    
  filecolor=black,
  linkcolor=black,
  urlcolor=black
}

\usepackage{textcomp}

\begin{document}

\begin{titlepage}
\begin{center}
\LARGE{\textbf{Measurement of the temperature distribution inside a calorimeter}}\\
\vspace{0.8cm}

\LARGE{\textbf{Ákos Sudár}}\\
\Large
Engineering Developer, Faculty of Mechanical Engineering, \\ Budapest University of Technology and Economics, \\
\vspace{0.5cm}
\centering
\begin{figure}[h!]
        \centering
        \begin{minipage}{5cm}
        \centering
        \includegraphics[width=5cm]{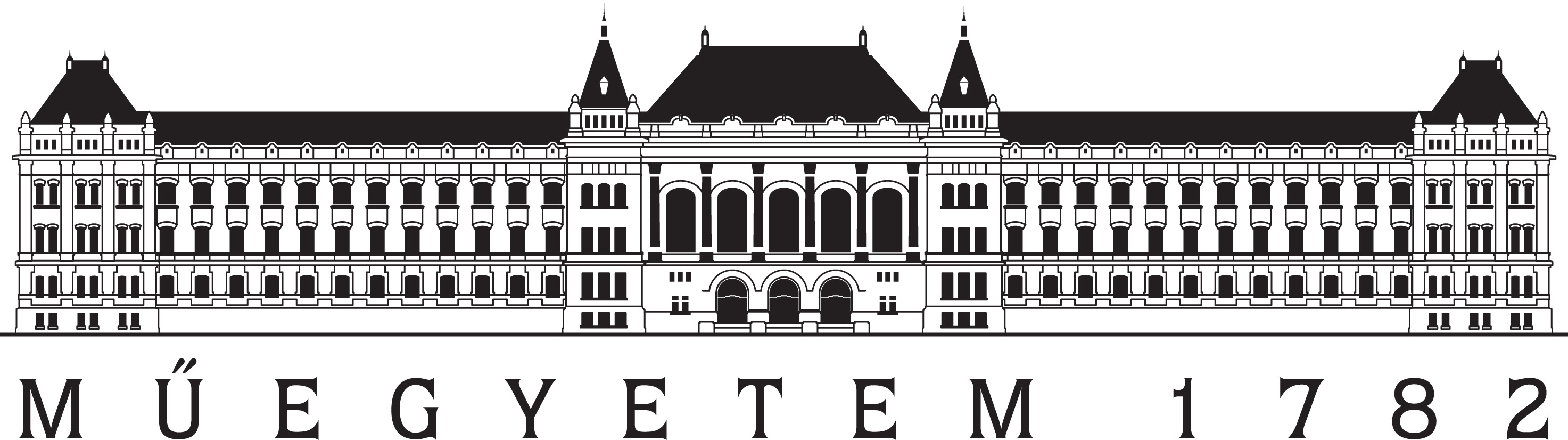}
        \end{minipage}
        \qquad
        \begin{minipage}{4cm}
        \centering
        \includegraphics[width=2.5cm]{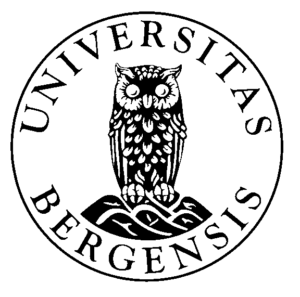}
        \end{minipage}
        \qquad
        \begin{minipage}{5cm}
        \centering
        \includegraphics[width=5cm]{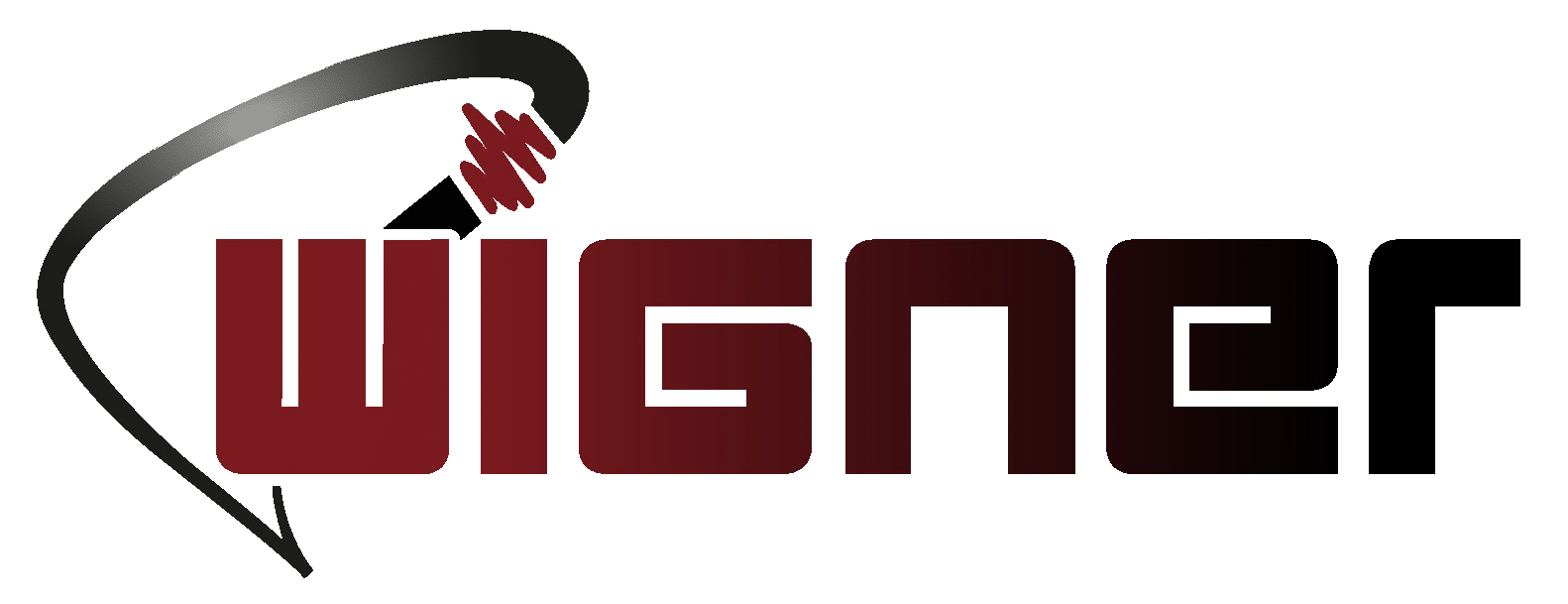}
        \end{minipage}
\end{figure}
\normalsize
\emph{Supervisor:}\\
Dr. Róbert Kovács, Ph.D.\\
Assistant professor, Department of Energy Engineering, \\ Budapest University of Technology and Economics, \\
Department of Theoretical Physics, Institute for Particle and Nuclear Physics, Wigner Research Centre for Physics\\

\vspace{0.2cm}
\emph{Consultants:} \\
Researcher, Mónika Varga-Kőfaragó, Ph.D. \\
RMI High Energy Experimental Particle and Heavy Ion Physics, \\ Wigner Research Centre for Physics\\

\vspace{0.2cm}
Dieter Røhrich, Ph.D. \\
Professor, Department of Physics and Technology, \\ University of Bergen \\

\vspace{0.2cm}
\large
Budapest, 2019

\end{center}
\end{titlepage}

\begin{titlepage}
\section*{Abstract}

Hadron therapy is a novel treatment against cancer. The main advantage
of this therapy causes less side effect in comparison to X-ray
irradiation methods. Hadron therapy is just ahead of a significant
breakthrough since this technique can be more precise, applying proton
computer tomograph (pCT) to map the stopping power in the tissues.

The research and development of a pCT require a fast detector to
measure the energy of hadrons behind the patient. The best detector
option is called hadron-tracking calorimeter, which consists of sandwich
layers of silicon tracking detectors and absorber layers. The
combination of measuring the trajectory (tracking process), and, in parallel, the
energy of relativistic particles, can provide high-resolution hadron
imaging. This semiconductor-based technology requires stable
temperature and homogeneous cooling.

I have worked in the development of this detector in the Bergen pCT Collaboration for two years. Last year my work was to investigate the temperature distribution in the
calorimeter and examine two cooling concepts in detail. I performed
both analytical and numerical calculations to analyze the temperature
distribution of the calorimeter. The final decision about the design takes into account
many engineering aspects, such as reliability, flexibility, and
performance. 
\end{titlepage}

\tableofcontents

\newpage
\section{Introduction}
\subsection{Cancer}

Cancer has become one of the leading reasons for death in the developed world. It is responsible for 25\% of all death in Hungary \cite{Cancerstatistics}. It affects all age groups, but the risk of cancer is increasing with age, as one can see in Figure \ref{fig:procure}.

\begin{figure}[H]
\centering
\begin{minipage}{16cm}
\centering
\includegraphics[width=16cm]{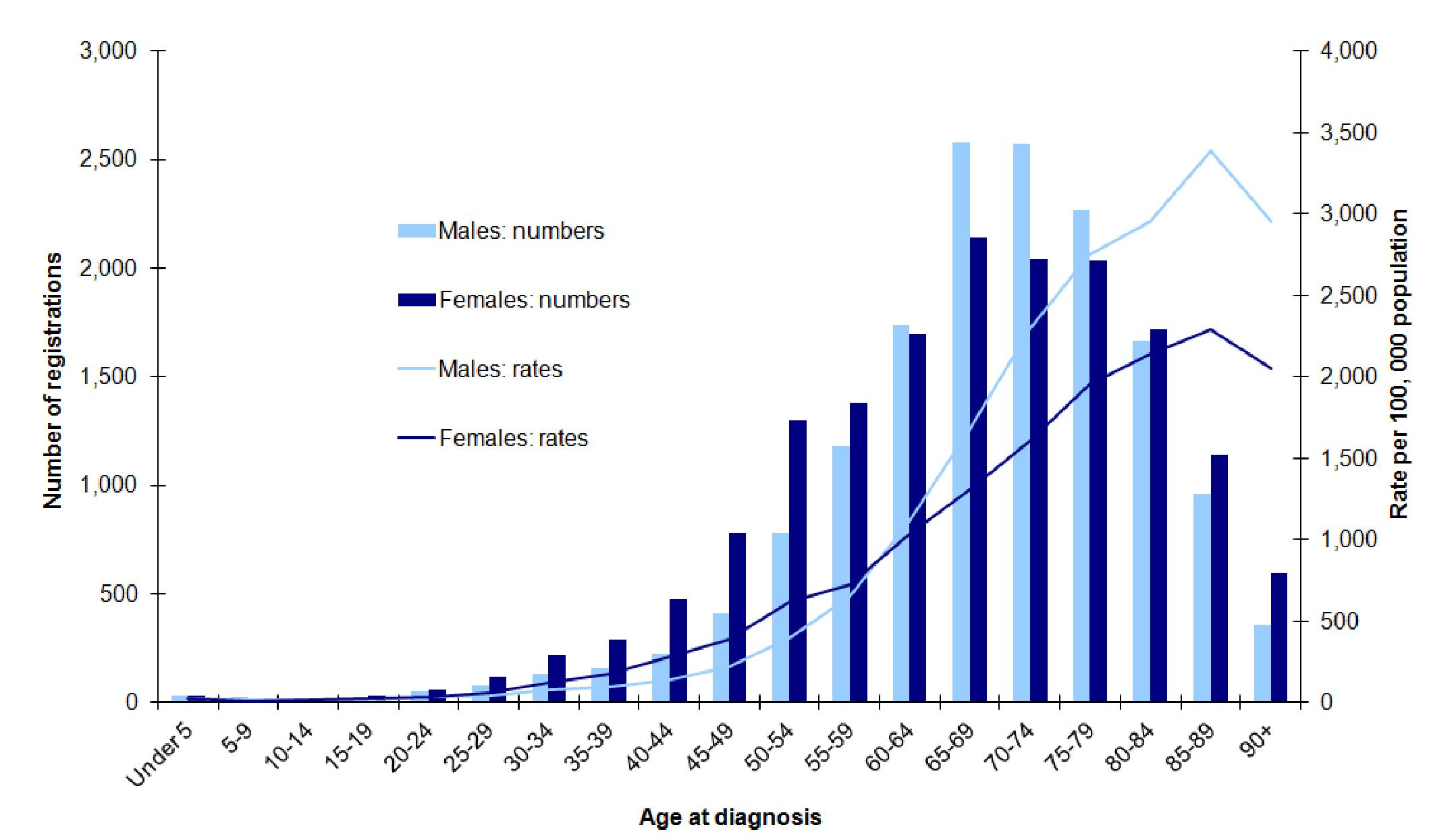}
\caption{Cancer affects all age groups, but the risk of cancer is increasing with age. This statistic was made in Scotland in 2016 \cite{cancerage}.}
\label{fig:cancerstatistics}
\end{minipage}
\end{figure}

This section is about the nature of cancer, briefly summarized based on the literature \cite{oncology1, oncology2, oncology3}.

The cell cycle is an important period of life of the cells because they divide and replace dead cells. In healthy tissues, there are stem cells, which can divide at any time. However, for most of the cells, the number of divisions and their times are limited. This is not true for the cancer cells because a DNA mutation turns off this limitation.

Usually, a cell dies after a lifetime, or in case of serious injury in its DNA. This function is called apoptosis, which, in general, does not work in cancer cells.

There are two types of cancer. The first is called bengilus. In the case of bengilus cancer, the tumor does not reduce the life expectancy and quality of life of the patient. The second one is called malignus, which means risk in life or the life quality of the patient. It is hard to determinate whether a tumor is bengilus or malignus. Usually, bengilus tumors expand slower, and they have a tegument, but the lack of these effects does not evidence that the tumor is malignus.

Unfortunately, tumors can expand in the surrounding tissues. This phenomenon is called an invasion. Cells of the tumor can spread far from the original one, this is called metastasis. After the treatment of cancer, it is still possible to find other cancer cells in the surrounding tissues that can cause a new, so-called residual tumor.

There are status groups, called stages, to categorize the tumors. The traditional staging has five stages:
\\0. stage: there is no invasion,
\\I. stage: small, local tumor, with small invasion (without metastasis),
\\II. stage: local tumor with more significant invasion, probably with metastasis in the local  lymph nodes,
\\III. stage: local tumor with significant invasion with high probability to has big metastasis in the local  lymph nodes,
\\IV. stage: local tumor with huge invasion and with far-reaching metastasis.

\subsection{Mapping methodes in the detection and treatment of cancer}

There are lots of methods to measure or find a tumor in the human body. 
This subsection presents them briefly based on \cite{oncology4}.

Ultrasound method uses sound waves to take a picture from the inside of the body. Applying this method, the sound wave penetrates into the body of the patient and is scattered on the boundaries of the organs with different speed of sound. A microphone records that reflection and a computer reconstructs the map.  A usual ultrasound machine takes a 2D picture from a segment of the human body, but nowadays it is also possible to obtain a 3D picture. The ultrasound technology does not have any side effect.  

An X-ray computer tomograph uses X-ray beams to fluoroscope through the human body. It creates a 3D picture from the nucleon density of the body. This type of imaging causes radiation in the body, which can damage healthy tissues.

The MR technology is based on the measurement of the magnetic field caused by the water molecules in the human body. The water molecules are excited by a strong magnetic field generated by superconductor magnets. The detector measures the changes of the magnetic field after the magnets are turned off. For instance, it makes visible the blood flow in the veins. MR technology does not cause any side effect, based on the current knowledge of science. The mapping process takes a long time, and the patient have to lie in a narrow, and closed space that causes feeling of being locked up in $4-6\%$ of the patients.

Mapping with radiative isotopes can give information about the life cycle of a tumor. This technology is based on the density of the chemical material which is absorbed by the tissues and the tumor. Usually, the tumor absorbes different amount of chemical element than the surrounding healty tissues. Thus, the concentration difference of the isotopes is measured here. The directional dependence of the radiation makes possible to construct the 3D map. Moreover, the amount of the absorbed radiation offers information about the type, size and the current state of the tumor. It is also useful to find metastasis; however, the irradiation of healthy tissues is unavoidable.

\subsection{Treatment of cancer}

The treatment aims to ensure high life expectancy (or as high as possible) and life quality (or as high as possible). In order to reach this goal, modern health care uses surgery, radiotherapy, chemotherapy, and hormone therapy. These treatments can be applied either standalone or combined as a complex therapy to improve the effectiveness.
That subsection is summarized based on \cite{oncology5, oncology6, oncology7}.

Every treatment starts with canceling the tumor.  It is crucial not to leave any cancer cell in the surrounding tissues because it can cause a residual tumor later. If it is not possible due to some reason, then it becomes necessary to kill these cells in the second part of the treatment. That part is about to kill all the remaining cancer cells in order to avoid metastasis. These cells are probably to be a metastasis, or cells in the vascular and lymphatic system, which try to find a new living space and create a new metastasis.

Surgery is the leading treatment of cancer. It is the oldest treatment of tumor diseases. In general, it is used to remove the original tumor, but it is useful in removing the metastasis, too. Its one of the most significant advantage is that the cancer cells cannot become resistive to the surgery. During this process, doctors cut out the tumor and usually remove the local lymph nodes also from the body to reduce the risk of metastasis. The surgery burdens the body of the patient. If the patient is not in adequate conditions, the operation becomes impossible. There are other situations in which the outcome is the same: either the tumor is in a hardly available location, or too big to cut out. In these cases, doctors have to find another type of treatment or use complex methods.

Radiotherapy uses the effects of the ionization of X-ray or ion beams. The beam exerts its effect in two ways, directly and indirectly. The direct way is when the particles of the beam damage the DNA and the critical parts of the cell. The indirect way is when the particles of the beam ionize the water molecules, and these molecules damage the DNA of the cancer cells. In both cases, the damage induces a self-destruct function of the cancer cell, thus the cell destroys itself. This self-destruction process is called apoptosis. Unfortunately, there are such cancer cells which have resistance to the radiation, the apoptosis process is prevented. 

Usually, chemotherapy uses apoptosis to destroy cancer cells. In the case of chemotherapy, doctors give medicine to the patient, which consists of chemical elements, inducing the apoptosis process. The most common chemicals are called cytostatics. 
The cytostatics affect the cancer cells more seriously than healthy cells. Besides, there are other chemicals to reduce the effect of cytostatics in the healthy regions or increase their exterminating effect in the cancer cells.

Surgery as a standalone treatment is usual for a tumor in the I.~or II.~stages. In the III.~stage, radiotherapy is recommended. Chemotherapy is typical for IV.~stage. 

Beyond these possibilities, sometimes the surgeon must remove the entire organ, this is called `organ removal surgery'. However, this is not preferred and tried to be avoided. For instance, using radiotherapy first reduces the size of the tumor, hence it is possible to preserve the organ as much as possible.
Moreover, chemotherapy helps to reduce the risk of metastasis with canceling the remaining cells in the vascular and lymphatic system.

\subsection{Radiotherapy}

Radiotherapy uses the ionizing effect of radiation to destroy cancer cells. There are different ways to transfer the radiation to the tumor \cite{OrvosiBiofizika}. As a first option, one can use beams, which penetrates the tissues and the tumor, and in the meantime, it damages them. This treatment is called external beam radiation therapy. Its advantage is that it is usable in many segments of the body, without using surgery. Secondly, one can situate a radioactive source in the vicinity of the tumor, for example, into a body cavity. This technology is called brachytherapy. The advantage is the same, no operation is needed. As a third option, it is possible to ingest certain radioactive isotopes into the body. In fact, a tumor absorbs more isotopes than any of the other healthy organs. This treatment is called systemic radiation therapy. The disadvantage of this technology is that the patient becomes radioactive for a short time, so he or she has to stay in the hospital for some days.

One of the primary development in radiotherapy is to reduce the side effects caused by the radiation in healthy tissues. If one improves the ratio of the ionization in the tumor, it reduces the amount of ionization in the healthy tissues, thus it reduces the side effects.

\begin{figure}[H]
\centering
\begin{minipage}{12cm}
\centering
\includegraphics[width=12cm]{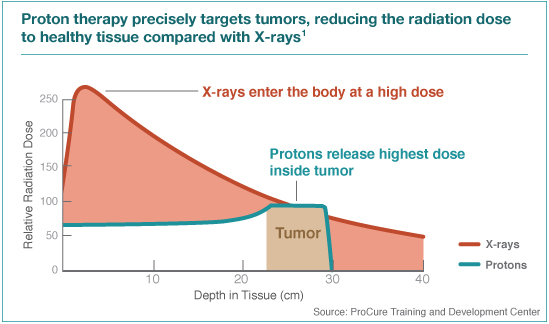}
\caption{The energy-loss of X-ray photons and protons in function of the amount of material in their previous path \cite{ProCure}.}
\label{fig:EnergiaLeadas}
\end{minipage}
\end{figure}

\begin{figure}[H]
\centering
\begin{minipage}{12cm}
\centering
\includegraphics[width=12cm]{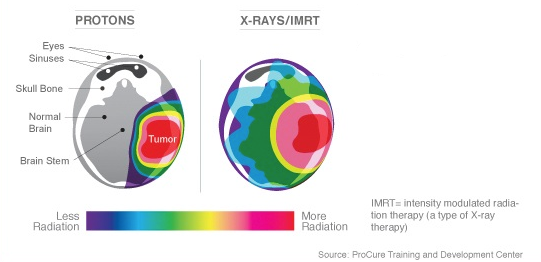}
\caption{Radiation in a head in case of proton and X-ray therapy \cite{ProCure}.}
\label{fig:procure}
\end{minipage}
\end{figure}

X-ray photon beams are the most common in external beam radiation therapy. The usage of protons or heavier ions instead of X-ray photons results in less ionization in the healthy tissues \cite{OrvosiBiofizika, protonTherapy0}. Its advantage is based on the radiation concentration distribution. The X-ray photons generate the highest radiation after the entry point in the body, then the radiation is monotonously decreasing. Using ions instead of X-ray photons, it is possible to concentrate the radiation onto the tumor, since the ions stop in the tumor mostly. Ions create the highest radiation right before they stop.
 One can see the resulted radiation by X-ray photons and protons in Figure \ref{fig:EnergiaLeadas}. One can also see the radiation of each treatment in a cross-section of a head in Figure \ref{fig:procure}.

\begin{figure}[H]
\centering
\begin{minipage}{12cm}
\centering
\includegraphics[width=12cm]{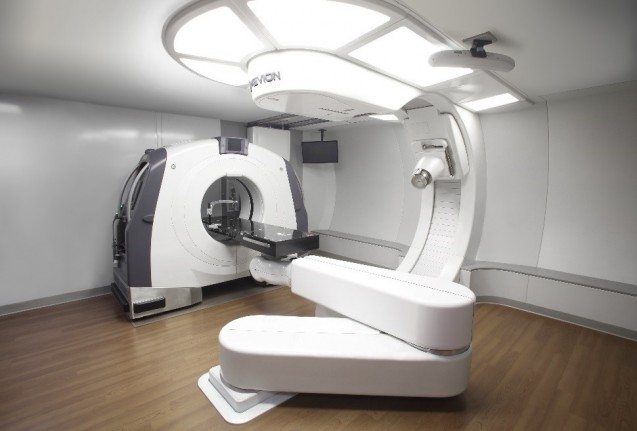}
\caption{The patient is fixed into a bed during the radiotherapy. The patient is moved with the bed by a robot arm that can be moved in all directions. First of all they take a 3D map of the tissues nearby the tumor of the patient, after they treat the tumor by a proton beam \cite{pKezeles}.}
\label{fig:ProtonKezeles}
\end{minipage}
\end{figure}

As it is apparent, the application of ions in the external beam radiation therapy is beneficial. Nowadays, the number of hadron therapy centers is increasing. One can see the treatment room of a hadron therapy facility in Figure \ref{fig:ProtonKezeles}. However, there is a design barrier in the spread of hadron therapy centers. This barrier is the lack of an accurate imaging tool for  dose planning. 
Doctors use X-ray computer tomographs (CTs) to take a three-dimensional map from the nucleon density of the body. However, the energy loss of ions depends on the electron density of the body. It is possible to calculate the electron density using the measured nucleon density, but there is no simple relation between them, because of the different ratio between proton and nucleon number of atoms of the human body. As a result, one can calculate the electron density with $1.7\%$ statistical error from nucleon density \cite{CTcomparison}. If one could measure the electron density map of the human body directly with the usage of ions, the statistical error could be reduced to $0.5\%$ without significant error \cite{CTcomparison}. Overall, the development of a hadron computer tomography (usually called proton computer tomography or pCT) is beneficial.

\subsection{Proton computer tomography}

The original idea of computer tomography (CT) comes from Allan M. Cormack (1963) \cite{Cormack}. He won the Nobel Prize with this idea in 1979.

\begin{figure}[H]
\centering
\begin{minipage}{12cm}
\centering
\includegraphics[width=12cm]{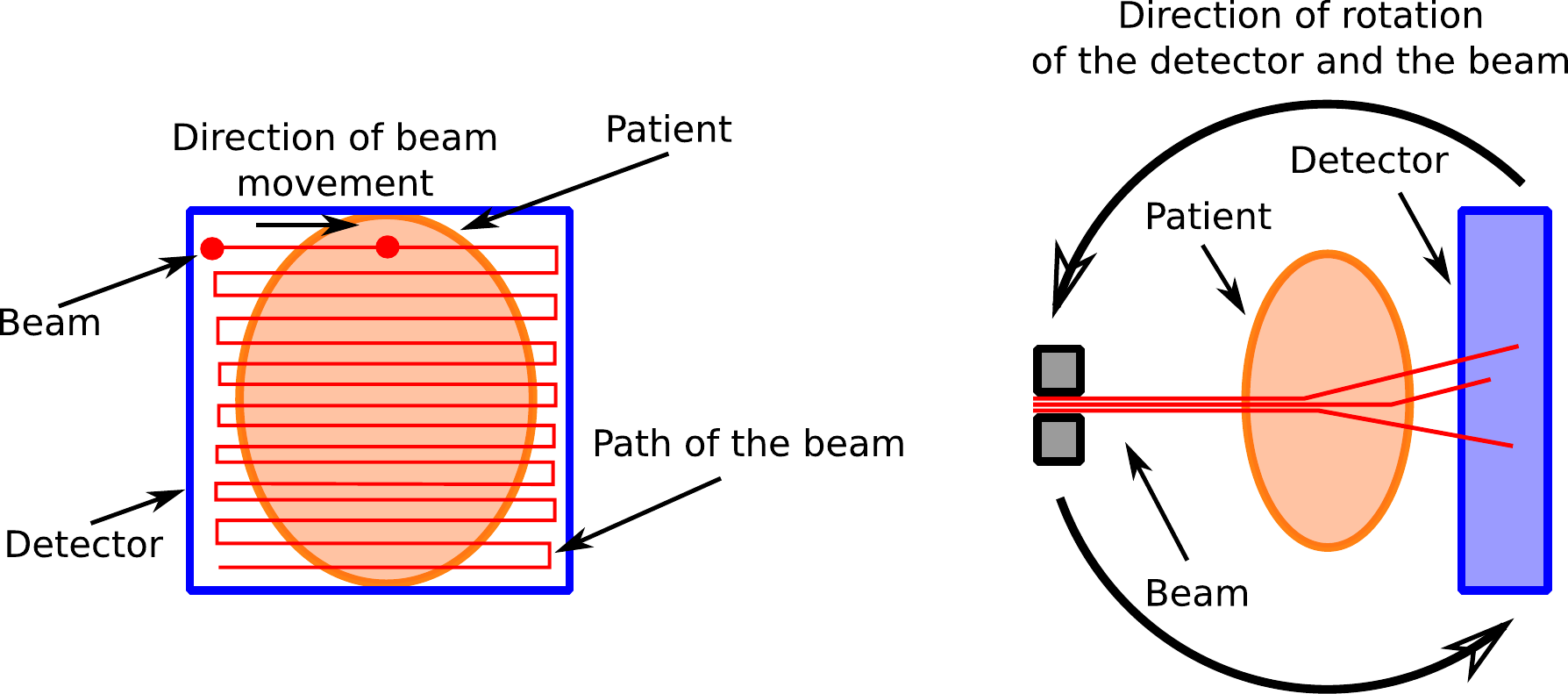}
\caption{Left: it is visible how the proton beam scans the body. Right: it is visible how the detector and the beam rotate around the patient. }
\label{fig:workingofapCT}
\end{minipage}
\end{figure}

A CT equipment moves around the body and takes pictures from the inside in several directions, and a computer calculates the 3D map using these pictures. In case of proton CT, one can use a beam to fluoroscope through the body. One scans the body with this beam in a particular direction, so one obtains a two-dimensional picture about the body. After, the detector is turned into a different direction, and the scanning process starts again. One can see the scanning and rotating process in Figure \ref{fig:workingofapCT}. Finally, a computer algorithm calculates the three-dimensional map of the body from the two-dimensional pictures.

During the scanning process, one has to use another detector at the other side of the patient in order to measure the energy and the path of particles coming from the beam. Such a detector is called a tracking calorimeter since it measures both the energy and the tracks of hadrons. $10^7-10^9~\frac{particle}{second}$ particle rate is necessary to obtain a three-dimensional picture within a reasonable time for clinical use \cite{Dieter}. To measure the path of the particles with such a rate was impossible five years ago. However, thanks to the development of tracking silicon pixel detectors in the last years in CERN LHC, there are available tracking detectors that can meet with this particle rate now \cite{ALICETEC}. These tracking detectors can measure the path of the particles, but there is not any available detector that can measure the energy of each particle with this particle rate.

\begin{figure}[H]
\centering
\begin{minipage}{12cm}
\centering
\includegraphics[width=12cm]{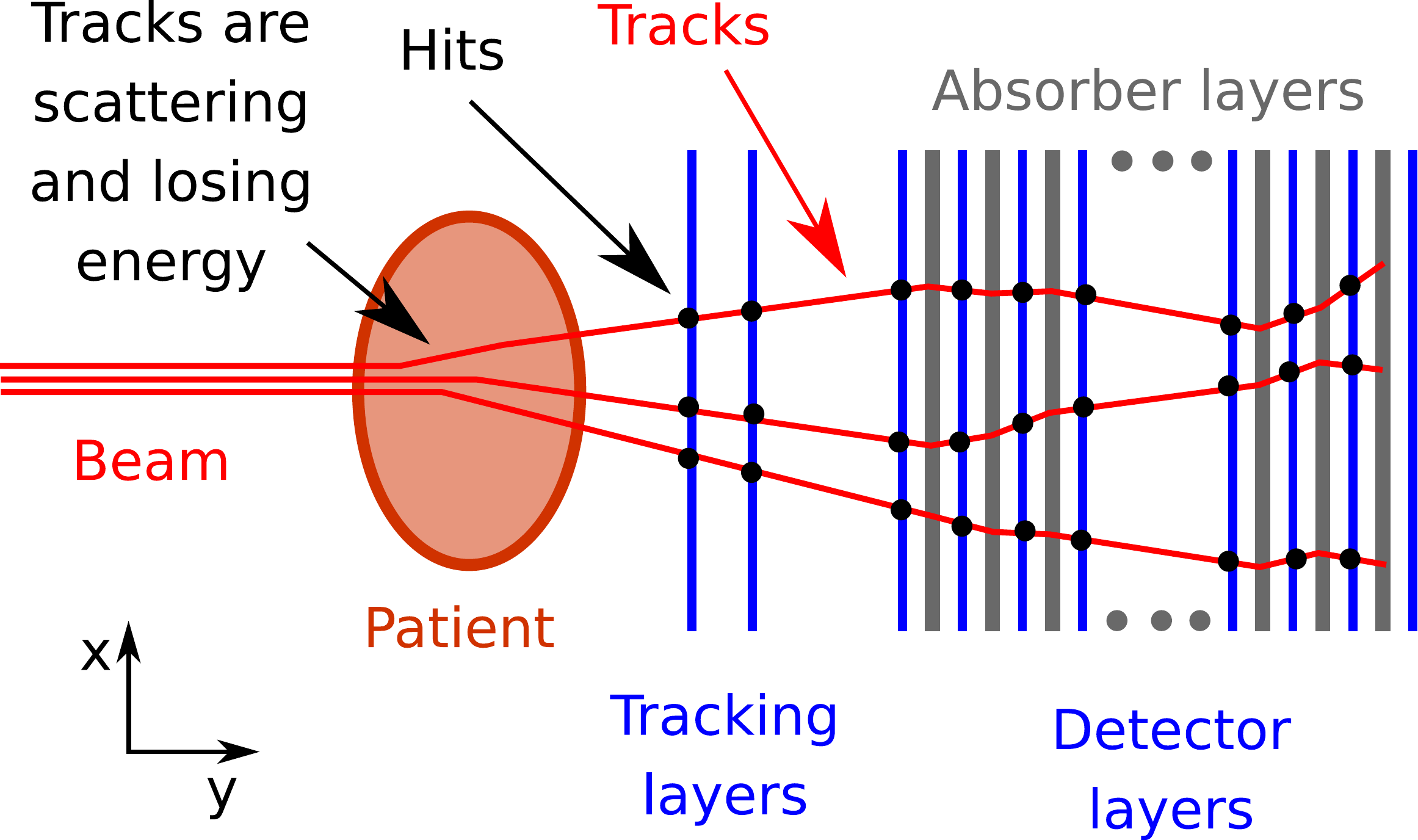}
\caption{Concept of the proton CT and its tracking calorimeter.}
\label{fig:Concept1}
\end{minipage}
\end{figure}

The Bergen pCT Collaboration aims to develop a calorimeter (calorimeter is a detector, which measures the energy of hadrons) build of alternating tracking detector and aluminum absorber layers, as a sandwich structure \cite{Dieter}. One can see the concept of this detector in Figure \ref{fig:Concept1}. As a tracking calorimeter, we are going to use a tracking detector, which was developed in ALICE, CERN LHC, and which is called ALPIDE.

My role in the Bergen pCT Collaboration is the investigation of the temperature distribution in the calorimeter and compare two cooling system concept. That work is crucial to ensure the accuracy of the detector. That part of the detector development is clearly separated from the other tasks related to the design. Thus the present thesis reflects my work.



\subsection{ALPIDE tracking detector}

\begin{figure}[!h]
\centering
\begin{minipage}{10cm}
\centering
\includegraphics[width=10cm]{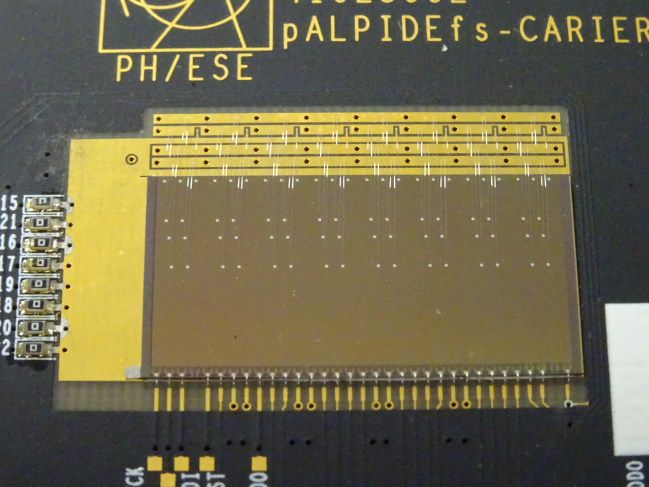}
\caption{ALPIDE detector \cite{MoniKonyve}.}
\label{fig:ALPIDE}
\end{minipage}
\end{figure}

ALPIDE was developed to upgrade the Inner Tracking System (ITS) of `A Large Ion Collider Experiment' (ALICE) in 2019-2020, which is one of the four large experiment programs of the Large Hadron Collider of the European Organization for Nuclear Research (CERN) 	\cite{ALICE_ALL}. One can see the ALPIDE detector in Figure \ref{fig:ALPIDE}.

\begin{figure}[!h]
\centering
\begin{minipage}{10cm}
\centering
\includegraphics[width=10cm]{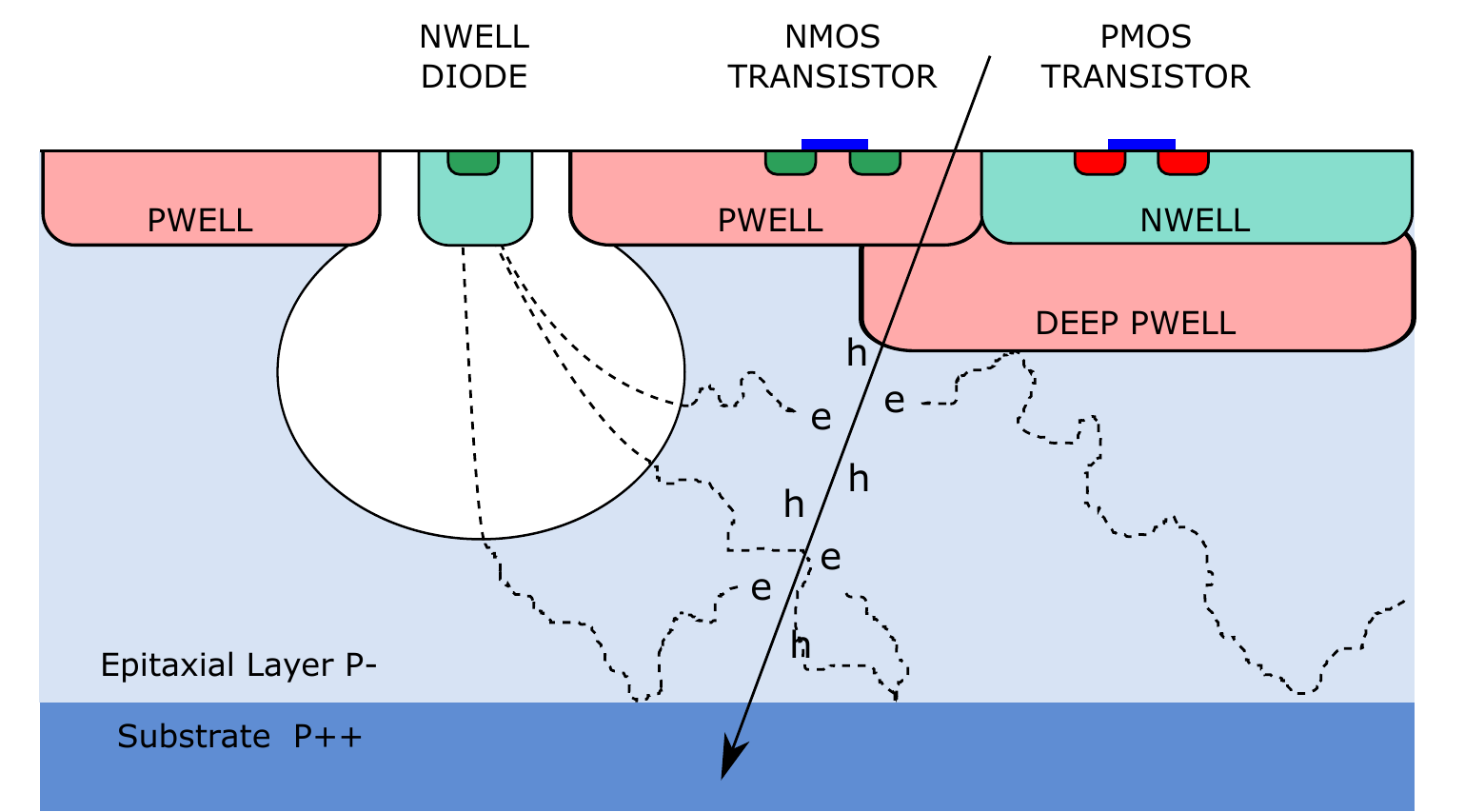}
\caption{Main parts of a monolithic active pixel sensor in a section view of one pixel \cite{ALICETEC}.}
\label{fig:MAPS}
\end{minipage}
\end{figure}

ALPIDE is a monolithic active pixel sensor (MAPS) type silicon detector \cite{ALICETEC}, which contains the sensitive sensors together with the readout electronics on the same silicon layer, see Figure \ref{fig:MAPS} for details. 
 When a charged particle goes through the sensitive (light blue area, called Epitaxial Layer P-) area of the detector, it generates electron-hole pairs. The electrons and the holes are transferred in the sensitive area by diffusion. The holes are absorbed by the Substrate P++ layer (dark blue). There is an area (white), which is evacuated by electric voltage. When an electron reaches this zone, the electric field of the evacuated zone transfers it into the NWELL diode (green). This diode collects the electrons, then a microelectronic amplifier amplifies the signal. Another microelectronics decides whether the signal is higher than a threshold or not. If so, the microerectronic device sends a signal to the readout electronics with the coordinates of the pixel. This is a state of the art solution because the sensitive silicon layer contains the complete amplifier electronics. It is possible, thanks to the DEEP PWELL (red), which separates the NWELL parts of the microelectronics, thus the NWELL parts cannot operate as an undesirable diode.

The electrons, generated by the same particle, can be transfered to more than one pixels. This effect causes that more than one pixel sends signals to the readout electronics. We call as cluster the pixels send signals after a particle go through the detector. The size of the cluster depends on the absorbed energy in the detector, so it is possible to use this information to increase the accuracy of the calorimeter. However the cluster size also depends on the temperature of the detector, so it is very important to know the accurate temperature of the detector, because a small difference can cause error in the energy measurement.

\newpage
\section{Detector}
\subsection{Concept of the detector}

\begin{figure}[!h]
\centering
\begin{minipage}{12cm}
\centering
\includegraphics[width=12cm]{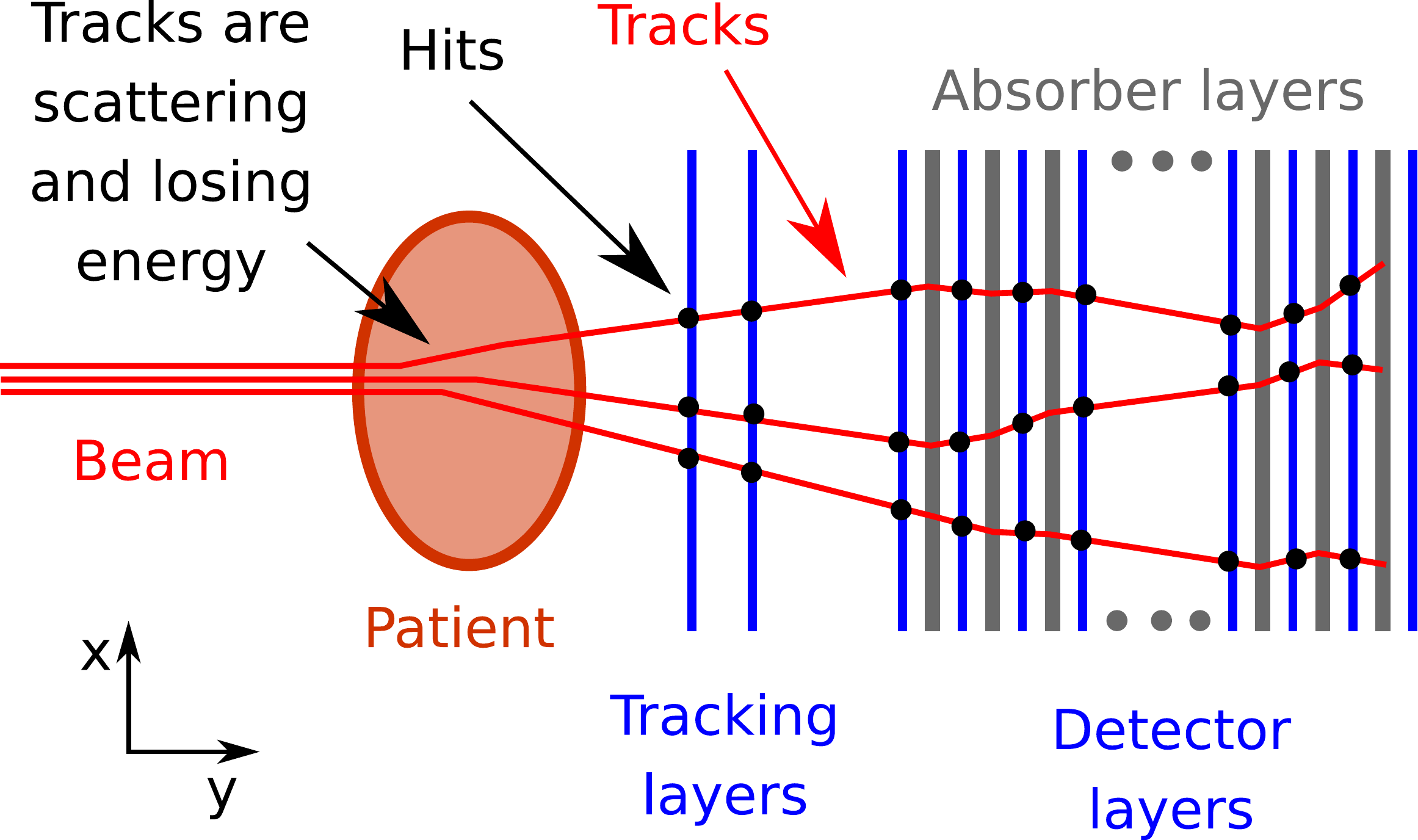}
\caption{Concept of the proton CT and its tracking calorimeter.}
\label{fig:Concept}
\end{minipage}
\end{figure}

Let us recall the concept of the detector in Figure \ref{fig:Concept}. The detector is built of two parts. The first part is a tracking detector, made of two layers of ALPIDE sensors. This part measures the direction of incoming particles. The second part of the detector is a calorimeter. It has a sandwich structure, and it is made of 35 ALPIDE sensor layers, separated by 4 mm thick aluminum absorber layers. This calorimeter measures the energy of the particles.

\begin{figure}[H]
\centering
\begin{minipage}{12cm}
\centering
\includegraphics[width=12cm]{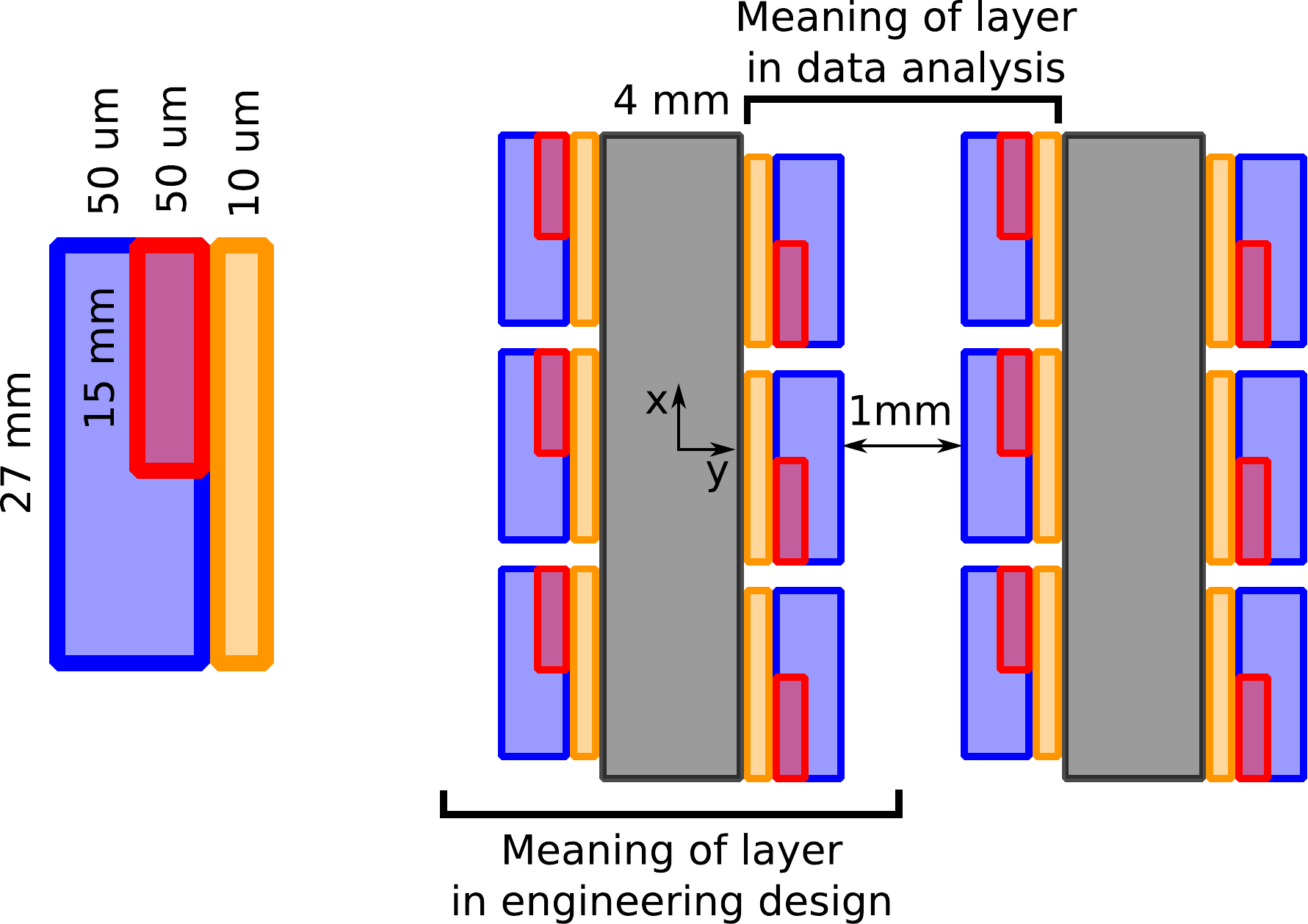}
\caption{This figure shows the structure of a layer. Gray: absorber, red: ALPIDE, blue: chip cable and orange: glue.}
\label{fig:Layers}
\end{minipage}
\end{figure}

One can see the structure of a detector layer in Figure \ref{fig:Layers}. The ALPIDEs are glued on both sides of the aluminum absorber layer, alternately with a small overlap, hence the absorber functions as a support structure also. Here, two definitions are introduced for the layers, depicted in Fig.~\ref{fig:Layers}. Its first meaning is about the engineering aspects, and consists of the aluminium absorber with APLIDE sensors on its both sides. Its second meaning is about the data analysis aspects, which considers the ALPIDE sensors between two aluminium absorbers. 


\begin{figure}[H]
\centering
\begin{minipage}{12cm}
\centering
\includegraphics[width=12cm]{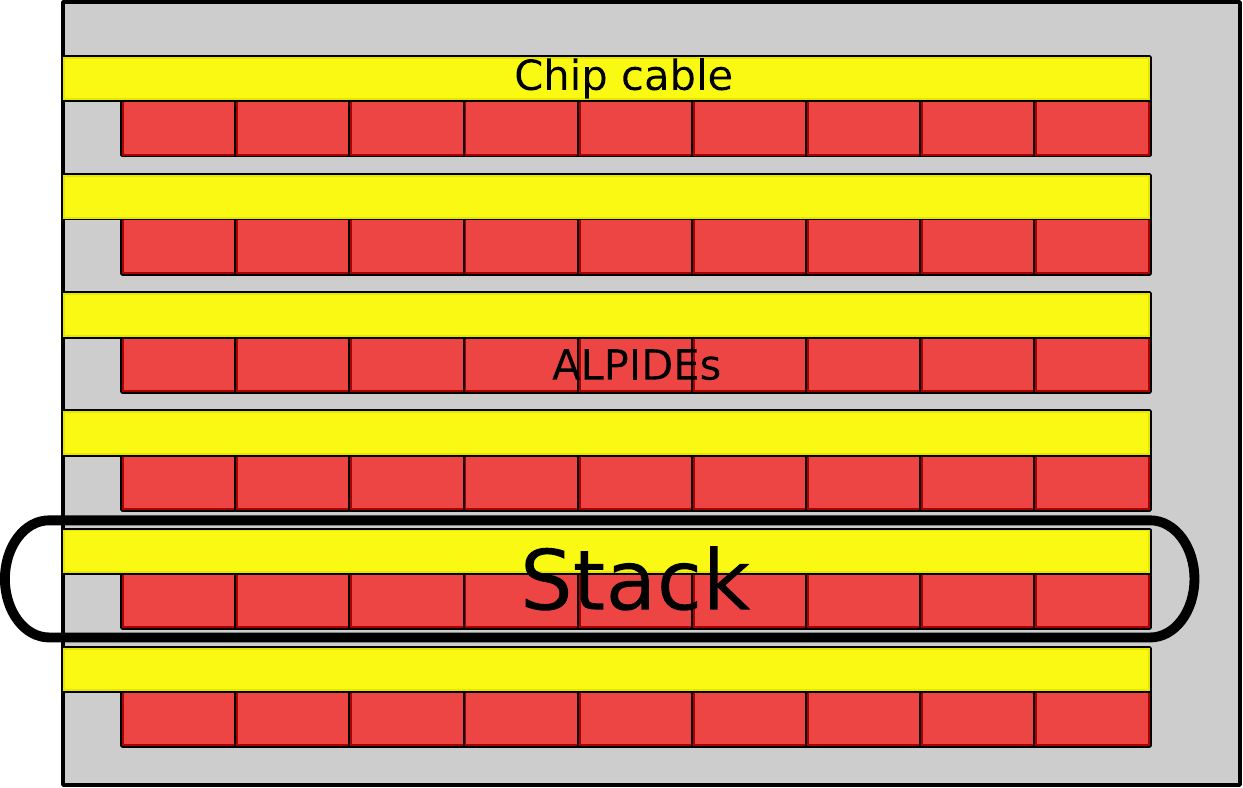}
\caption{A stack is a group of ALPIDEs, which is connected to the readout electronics with the same chip cable.}
\label{fig:Stacks}
\end{minipage}
\end{figure}

\begin{figure}[!h]
\centering
\begin{minipage}{12cm}
\centering
\includegraphics[width=12cm]{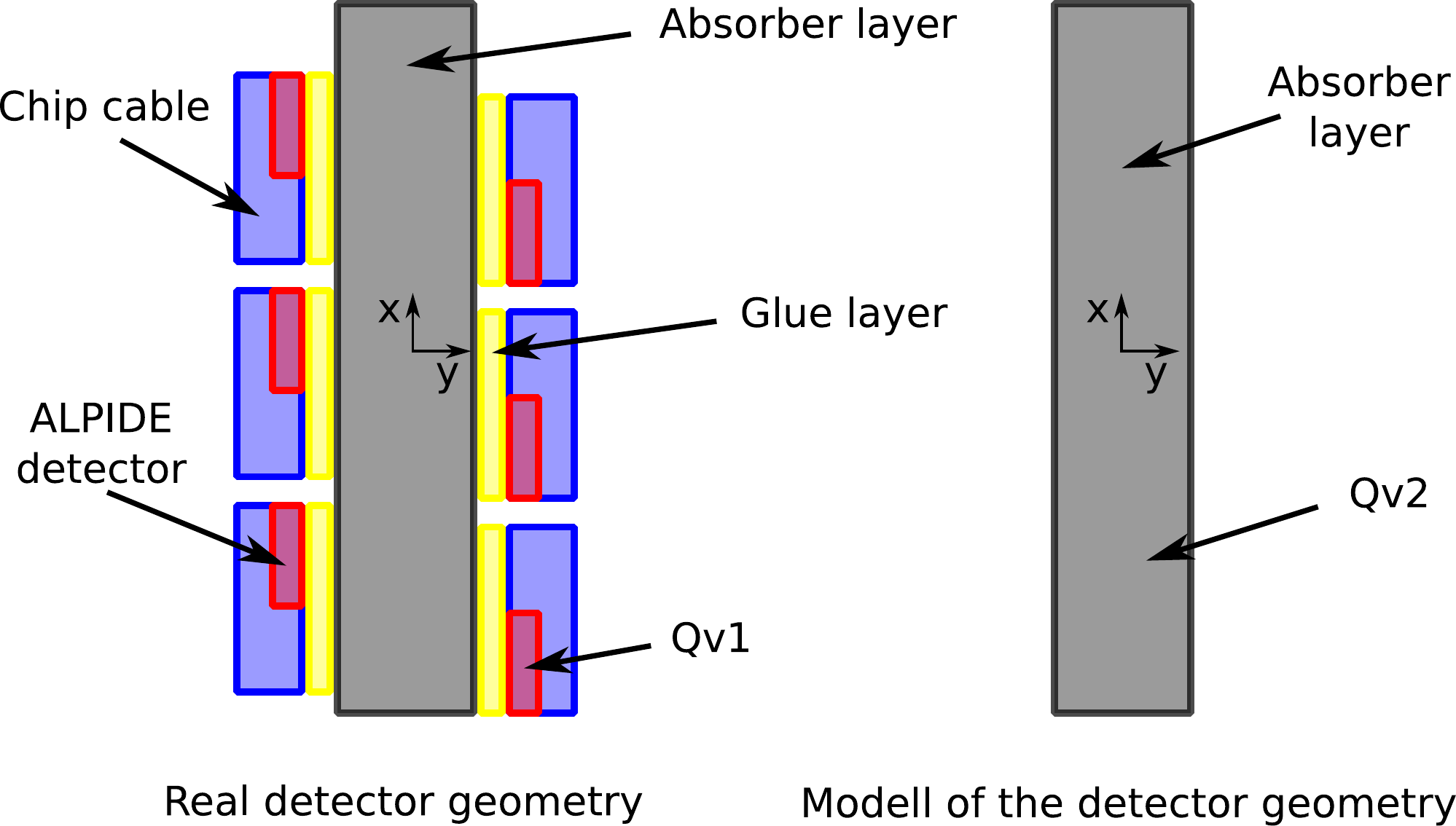}
\caption{Comparing the real detector to the model. $Q_{v_1}$ is the real volumetric heat generation in ALPIDEs, and $Q_{v_2}$ is the equivalent heat generation in absorber layer.}
\label{fig:Layer_Mosell}
\end{minipage}
\end{figure}

The ALPIDEs are grouped into stacks that form a readout unit. One can see a stack on one side of the absorber in Figure \ref{fig:Stacks}. The stacks are placed in this way, but they are mirrored.
The thickness of the glue, ALPIDE, and chip cable are small (Figure \ref{fig:Layers}) but not necessarily negligible. 
The resulted temperature difference $\Delta T$ in the perpendicular direction of the layer can be calculated with the following equation in steady-state \cite{hokozles}:

\begin{equation}
\Delta T=\frac{q \cdot h}{\lambda},
\end{equation}
where $q$ is the heat flux, $\lambda$ stands for the thermal conductivity, and $h$ being the thickness. One can find the constants and the results in Table \ref{tab:smalllayers}. Indeed, these $\Delta T$ values are small, thus in later calculations their thermal resistances are neglected. 
A layer model is depicted in Figure \ref{fig:Layer_Mosell}. The volumetric heat generation occuring in the ALPIDE detectors is modelled  as an equivalent volumetric heat generation in the absorber layer.

\begin{table}[h!]
  \begin{center}
    \begin{tabular}{|l|l|r|r|r|r|r|} 
      \hline
      \textbf{Layer} & \textbf{Material} & \textbf{Thermal} & \textbf{Thickness} &  \textbf{Typical} & \textbf{Temperature}\\
      \textbf{} & \textbf{} & \textbf{conductivity} & \textbf{}  & \textbf{heat flux} & \textbf{difference}\\
      \hline
       &  & $\lambda~\big [\frac{W}{mK}\big]$ & $h~[\mu m]$ & $q~\big[\frac{W}{m^2}\big]$ & $\Delta T~[K]$\\
      \hline
      Glue & Glue & 0.22 & 10 & 410 & 0.0186\\
      ALPIDE & Silicone & 149 & 50 & 410 & $1.38\times10^{-4}$\\
      Chip cable & Polyamide & 0.24 & 50 & 410& 0.0854\\
      \hline
    \end{tabular}
  \end{center}
  \caption{Thermal resistance parameters of ALPIDE, chip cable and glue.}
  \label{tab:smalllayers}
\end{table}

One can see the essential geometrical parameters and the thermal load of a real layer in Table \ref{tab:layerinreality}. It is not worth to model it in its original arrangement. Consequently, a simplified model is prepared using the parameters in Table \ref{tab:layermodell}. There are several layers side by side in which almost equal heat generation occurs. Thus, applying a periodic boundary condition, one can define a `unit' in which half of the air gap appears on both sides together with one layer. Between the layer and the air heat convection occurs. 

\begin{table}[h!]
  \begin{center}
    \begin{tabular}{|l|l|r|r|r|r|r|} 
      \hline
      \textbf{Part} & \textbf{Material} & \textbf{Height} & \textbf{Width} &  \textbf{Thickness} & \textbf{Heat generation}\\
      \hline
       &  & [mm] & $[mm]$ & $[mm]$ & $\big [\frac{W}{m^3}\big ]$\\
      \hline
      Absorber & Pure aluminum & 200 & 300 & 4 & -\\
      ALPIDE & Silicone & 15 & 30 & 0.05 & $8.2\cdot 10^6$\\
      Chip cable & Polyamide & 27 & 285 & 0.05& -\\
      Glue & Glue & 27 & 270 & 0.01 & -\\
      Air gap & Air & 200 & 300 & 1 & -\\
      \hline
    \end{tabular}
  \end{center}
  \caption{The essential parameters of the real layer.}
  \label{tab:layerinreality}
\end{table}

\begin{table}[H]
  \begin{center}
    \begin{tabular}{|l|l|r|r|r|r|r|} 
      \hline
      \textbf{Part} & \textbf{Material} & \textbf{Height} & \textbf{Width} &  \textbf{Thickness} & \textbf{Heat generation}\\
      \hline
       &  & $[mm]$ & $ [mm]$ & $[mm]$ & $\big [\frac{W}{m^3}\big]$\\
      \hline
      Absorber & Pure aluminum & 200 & 300 & 4 & $8.3\cdot 10^4$\\
      Air gap & Air & 200 & 300 & 1 & -\\
      \hline
    \end{tabular}
  \end{center}
  \caption{Overall dimensions and the heat load in the layer model.}
  \label{tab:layermodell}
\end{table}

\subsection{Without any active cooling system}

Here, I consider the detector that is covered by a casing, it protects the detector from external effects. As a first approximation, I neglect the inner temperature difference, and the entire detector is modeled as a lumped capacitance. Despite this simplification, it offers quantitative information about the characteristic heating of the device, without any active cooling system.
One can see the material data in Table \ref{tab:materialmodell}, which are used to characterize the detector. In Table \ref{tab:freeairmodell}, further parameters are summarized for modeling. Table \ref{tab:generatedheat} contains the number of ALPIDEs and the heat generation is specified for each part. The amount of heat generation is determined based on the description of ALPIDE elements, provided by the research group. One can see the heat generated by the particles of the beam in Table \ref{tab:particleheat}. The heat generated by the beam is several orders of magnitude smaller than the heat generated by the electronics, so I neglected it. Regarding the heat transfer coefficient, I assumed air at rest around the detector casing.

\begin{table}[H]
  \begin{center}
    \begin{tabular}{|l|r|r|r|r|} 
      \hline
      \textbf{Part} & \textbf{Density} & \textbf{Mass} & \textbf{Specific heat } &  \textbf{Heat capacity}\\
      \hline
       & $\big [\frac{kg}{m^3}\big ]$ & $[kg]$ & $\big [\frac{J}{kg\cdot K}\big ]$ & $\big [\frac{J}{K}\big ]$\\
      \hline
      One layer & 2850 & 0.684 & 434 & 297\\
      35 layers & 2850 & 23.94 & 434 & 10395\\
      \hline
    \end{tabular}
  \end{center}
  \caption{Thermal parameters for the lumped capacitance model.}
  \label{tab:materialmodell}
\end{table}

\begin{table}[H]
  \begin{center}
    \begin{tabular}{|r|r|r|r|r|} 
      \hline
      \textbf{Height} & \textbf{Width} &  \textbf{Thickness} & \textbf{Surface area} & \textbf{Heat transfer coefficient}\\
      \hline
      $[mm]$ & $[mm]$ & $[ mm]$ & $[m^2]$ & $\big[\frac{W}{m^2\cdot K}\big ]$\\
      \hline
      200 & 300 & 175 & $0.295$ & $ 5 $\\
      \hline
    \end{tabular}
  \end{center}
  \caption{The dimensions of the detectors, and the applied heat transfer coefficient. }
  \label{tab:freeairmodell}
\end{table}

\begin{table}[H]
  \begin{center}
    \begin{tabular}{|l|r|r|r|r|} 
      \hline
      & \textbf{ALPIDE} & \textbf{Stack} &  \textbf{Layer} & \textbf{Detector}\\
      \hline
      Number of ALPIDEs & 1 & 9 & 108 & 3780\\
      \hline
      Generated power & 185 mW & 1.66 W & 19.9 W & 697 W\\
      \hline
    \end{tabular}
  \end{center}
  \caption{The generated heat and the number of ALPIDEs.}
  \label{tab:generatedheat}
\end{table}

\begin{table}[H]
  \begin{center}
    \begin{tabular}{|r|r|r|} 
      \hline
      \textbf{Energy of a particles} & \textbf{Particle rate} &  \textbf{Generated power}\\
      \hline
      $[MeV]$ & $[\frac{particle}{s}]$ & $[mW]$\\
      \hline
      200 & $10^9$ &  35.2\\
      \hline
    \end{tabular}
  \end{center}
  \caption{The generated heat by the particles of the beam.}
  \label{tab:particleheat}
\end{table}

\begin{figure}[H]
\centering
\begin{minipage}{12cm}
\centering
\includegraphics[width=12cm]{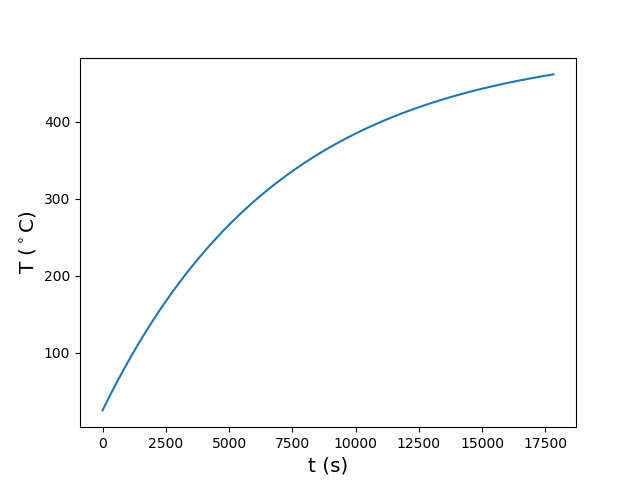}
\caption{The temperature history of the detector without active cooling.}
\label{fig:T_t_freeair}
\end{minipage}
\end{figure}

The lumped capacitance method is the mathematical formulation of the I.~law of thermodynamics, i.e., one has to formulate the balance of internal energy. On the left hand side, there is the time evolution of the internal energy, considering constant specific heat $c$ and mass density $\rho$. On the right hand side, I consider the thermal load and the heat transfer by convection \cite{hokozles}:
\begin{equation}
\label{eqn:powerballance}
\frac{dT}{dt}\cdot m\cdot c=Q-(T-T_{\infty})\cdot \alpha \cdot A~.
\end{equation}
The solution of the differential equation (\ref{eqn:powerballance}) reads:
\begin{equation}
\label{eqn:powerballancesolution}
T(t)=C_{const}\cdot e^{-\frac{ \alpha \cdot A}{m \cdot c}t}+T_{\infty}+\frac{Q}{\alpha \cdot A},
\end{equation}
in which I can define the time constant:
\begin{equation}
\label{eqn:timeconstant}
\tau=\frac{m \cdot c}{\alpha \cdot A}~,
\end{equation}
and the $C_{const}$ is about to consider the initial condition
\begin{equation}
T(0)=T_0.
\end{equation}
One obtains the following form of the $T(t)$ function  \eqref{eqn:powerballancesolution}:
\begin{equation}
T(t)=\frac{Q}{\alpha \cdot A} + T_{\infty}+(T_0 - T_{\infty} - \frac{Q}{\alpha \cdot A}) \cdot e^{\frac{-t}{\tau}}~.
\end{equation}
The maximum allowable temperature for the detector is $40~^\circ C$, which can be exceeded easily in continous operation, even for a one hour interval (see  Figure \ref{fig:T_t_freeair}), and Table \ref{tab:calsulationdata} summarizes the constants used in the calculation. The allowable $40~^\circ C$ exceeded under $230$ s, which time interval is too short for the detection. Moreover, one should wait too much time to use the detector again. Consequently, an active cooling system is desired.

\begin{table}[h!]
  \begin{center}
    \begin{tabular}{|r|r|r|r|r|r|} 
      \hline
      \textbf{$T_0$} & \textbf{$T_{\infty}$} &  \textbf{$Q$} & \textbf{$\alpha \cdot A$} & \textbf{$m \cdot c$} & \textbf{$\tau$}\\
      \hline
      [$^\circ C$] & [$^\circ C$] & $W$ & $\big [\frac{W}{K}\big ]$ & $\big[\frac{kJ}{K}\big]$ & $[s]$\\
      \hline
      25 & 25 & 697 & 1.48 & 10.4 & 7030\\
      \hline
    \end{tabular}
  \end{center}
  \caption{The constants used to determine the temperature history.}
  \label{tab:calsulationdata}
\end{table}

\subsection{Cooling concepts}
Here, I compare three significantly different cooling concepts:
\begin{itemize}
\item water cooling at the edge of aluminum layers,
\item air cooling between the layers,
\item water cooling inside the aluminum layers.
\end{itemize}
Two of them use water as coolant and one concept uses air. Let me review them in the following.

\begin{figure}[H]
\centering
\begin{minipage}{12cm}
\centering
\includegraphics[width=12cm]{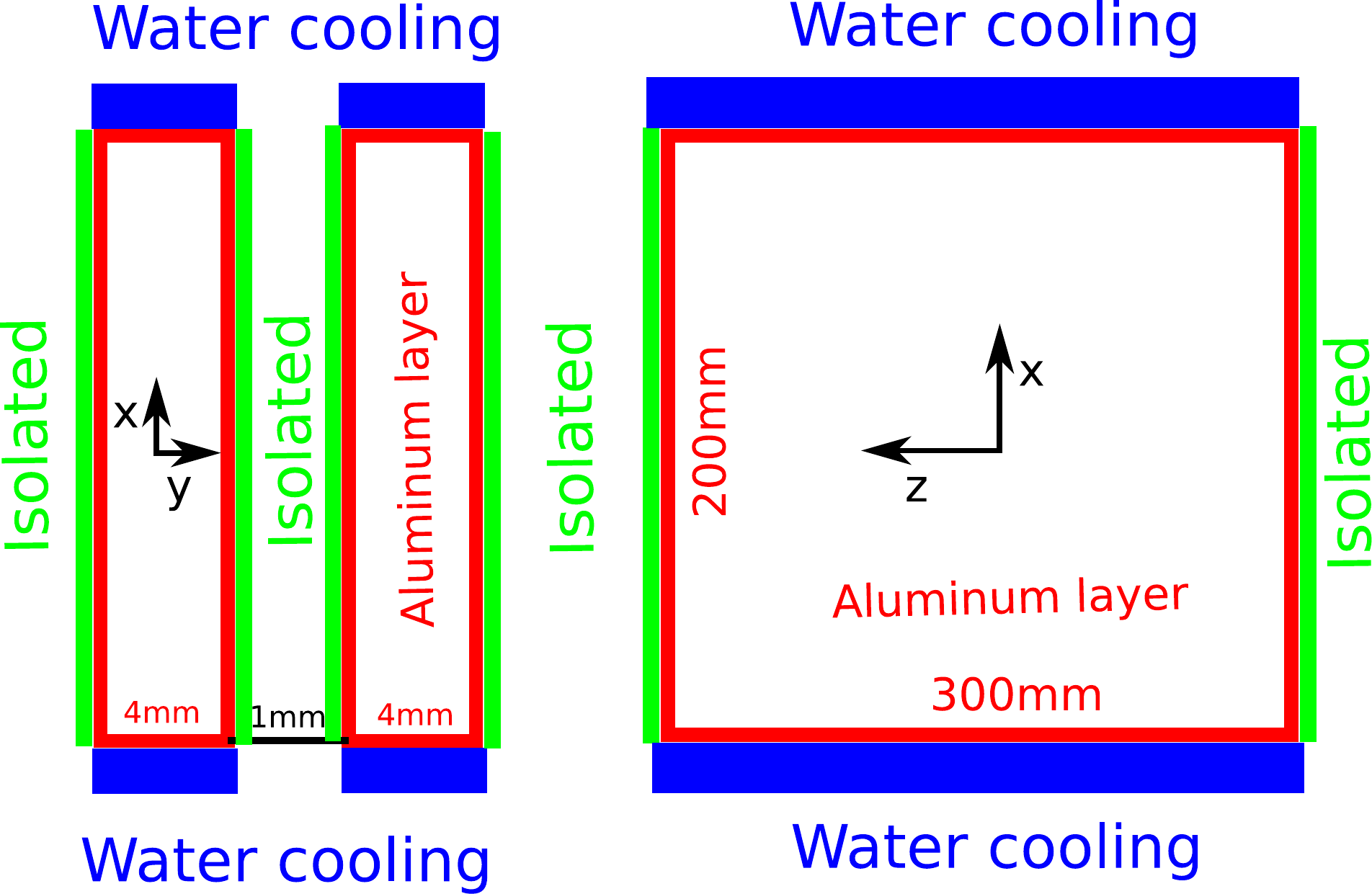}
\caption{Schematic arrangement of concept A.}
\label{fig:Water_Geom}
\end{minipage}
\end{figure}

\textbf{A) Water cooling at the edge of aluminum layers.}
In concept A, the detector layers are cooled at their top and bottom edges. The scheme of this concept is depicted in Figure \ref{fig:Water_Geom}. The benefit, compared with concept C, is that it does not need any extra material between the layers, thus it does not cause any extra scattering, consequently, it does not have any negative effect in the data reconstruction.

On one hand, this is a compact and quiet arrangement, so it is perfect to use in treatment as well. It does not need a high coolant volume flow because of the high density and heat capacity of water. One needs a heat pump to cool the coolant that can be set far from the detector, and it makes possible to change the mean temperature of the detector. On the other hand, we cannot influence the temperature distribution. That could be a problem when a significant temperature difference occurs in a layer. In this case, one cannot use this concept of detector cooling. It reveals that the knowledge about the temperature distribution could be essential.

\begin{figure}[H]
\centering
\begin{minipage}{12cm}
\centering
\includegraphics[width=12cm]{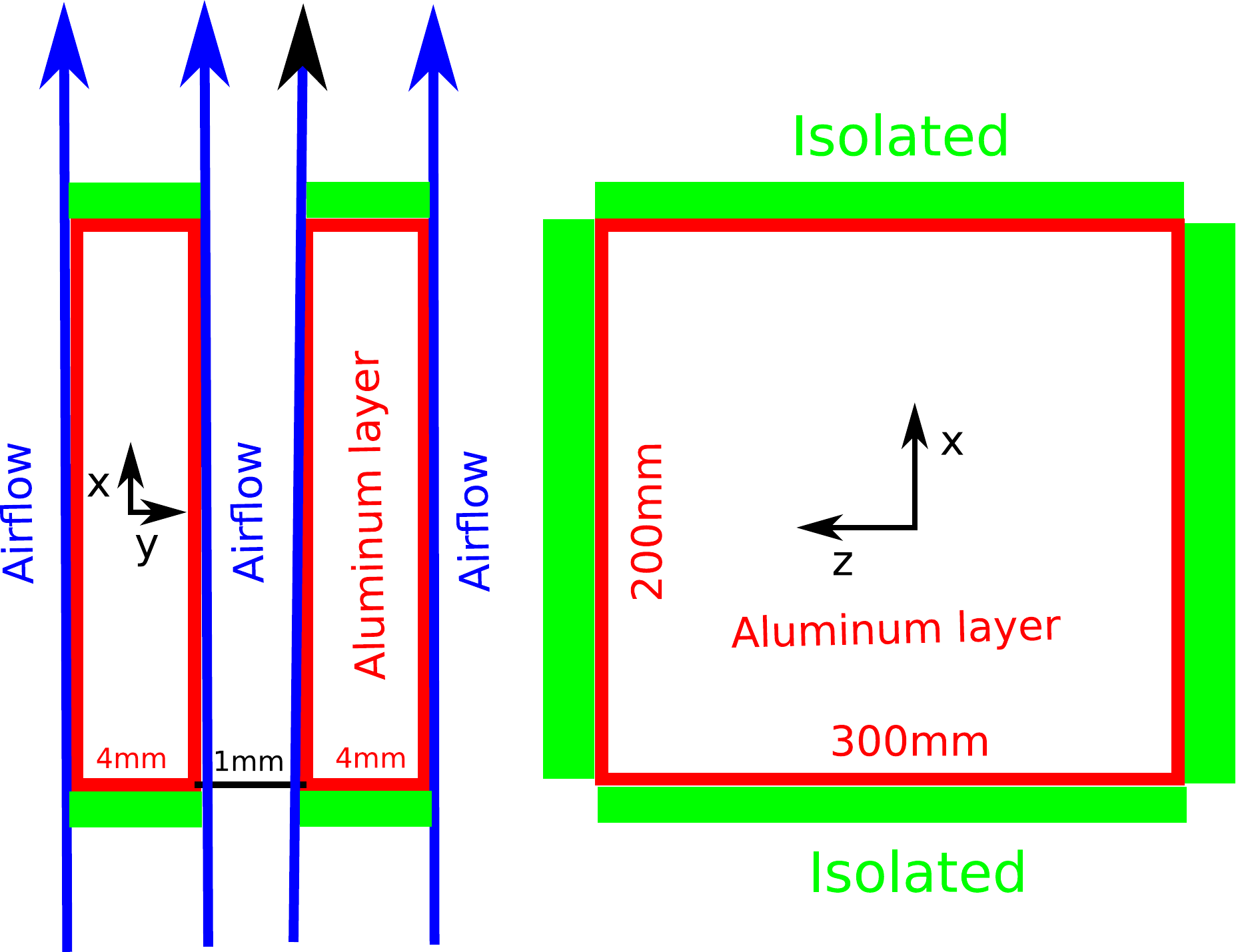}
\caption{Schematic arrangement of concept B.}
\label{fig:Air_Geom}
\end{minipage}
\end{figure}

\textbf{B) Air cooling between layers.} In concept B, there is forced airflow between the detector layers, which cools the detectors directly. One can see this cooling concept in Figure \ref{fig:Air_Geom}. Using this cooling method, almost the entire surface of the detector is exposed to forced convection. Here, I expect a more advantageous temperature distribution with less significant temperature gradient within a layer. 

It could be easier to realize this concept since one can use a simple fan instead of a water circuit system, thus the engineering design is easier than the others. Moreover, it becomes completely irrelevant about how to protect the electronics from the coolant. However, it uses the air of the room as a coolant which restricts the minimum temperature and the cooling rate. It may result in high airspeed and disadvantageous vibrations in the structure. It is also an interesting question to investigate the temperature distribution in this case. 

\textbf{C) Water cooling inside aluminum layers.}
This concept can produce the most homogenous temperature distribution. However, it is disadvantageous for the data analysis due to the flowing medium among the layer. Moreover, it would be significantly harder to protect the electronics in case of any failure. Later, this concept is excluded from further investigations.


\newpage
\section{Comparing the concepts}
Now, my aim is to present the basic differences between concepts A and B by comparing the resulting temperature distributions to each other. In the following, I consider only one-dimensional heat conduction along the layer (direction of x axis). In case of concept A I also used symmetry boundary condition at the middle ($x=0$) of the layer.

\subsection{Concept A}

\begin{figure}[!h]
\centering
\centering
\includegraphics[width=10.5cm]{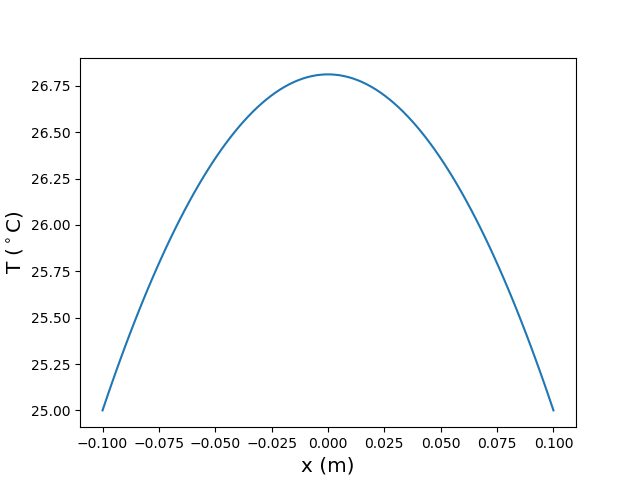}
\caption{Temperature distribution in the layer, where the direction of x axis is visible in Figure \ref{fig:Water_Geom}.}
\label{fig:Tdist_water}
\end{figure}

In general, the heat transfer coefficient can be high enough to replace the convection boundary condition with a so-called first-type boundary in which the temperature is prescribed directly. This is what we apply here. 
Integrating the Fourier heat equation in Cartesian coordinate system, one obtains \cite{hokozles}
\begin{equation}
T(x)= - \frac{q_v}{2 \lambda} x^2 + \frac{q_v}{2 \lambda} \left(\frac{L_1}{2}\right)^2 + T_0,
\end{equation}
in which $T_0$ is the coolant temperature that used as a boundary condition, and $L_1$ stands for the height of the detector.
Due to the symmetrical arrangement, the maximum temperature occurs at the middle, i.e., considering $x=0$,
\begin{equation}
T_{max}=\frac{q_v}{2 \lambda} \left(\frac{L_1}{2}\right)^2 + T_0.
\end{equation}
The desired operation condition requires as small as possible temperature difference inside a layer, thus the difference between $T_{max}$ and $T_0$ has the importance:
\begin{equation}
T_{diff}=T_{max}-T_0=\frac{q_v}{2 \lambda} \left(\frac{L_1}{2}\right)^2.
\end{equation}
One can see the temperature distribution in Figure \ref{fig:Tdist_water}, and the parameters can be found in Table \ref{tab:constantsofwatercooling}.

\begin{table}[!h]
  \begin{center}
    \begin{tabular}{|l|r|r|} 
      \hline
      \textbf{Constant} & \textbf{Notation} &  \textbf{Value}\\
      \hline
      Coolant temperature & $T_0$ & $25~^\circ C$\\
      Thermal conductivity of the absorber & $\lambda$ & $237~\frac{W}{m \cdot K}$\\
      Height of the detector & $L_1$ & $0.2~m$\\
      Equivalent volumetric heat generation in the absorber & $q_v$ & $8.3 \cdot 10^4~\frac{W}{m^3}$\\
      \hline
    \end{tabular}
  \end{center}
  \caption{Constants used to the calculation of the temperature distribution in concept A.}
  \label{tab:constantsofwatercooling}
\end{table}

\begin{figure}[H]
\centering
\centering
\includegraphics[width=10.5cm]{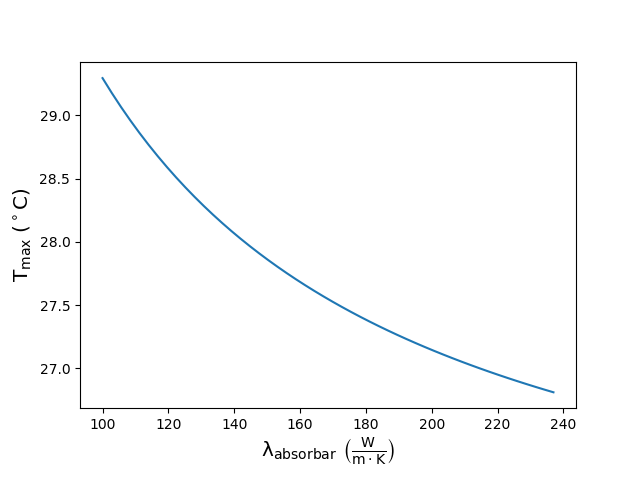}
\caption{Maximum layer temperature as a function of thermal conductivity of the absorber layer. The temperature of the edge of the detector is set to 25 $^\circ C$.}
\label{fig:Tmax_TC_Absorber_Water}
\end{figure}

In the current state of the detector design, the research group did not decide the material of the absorber. Most probably, it will be made from an aluminium alloy. Previously, the material properties of pure aluminium is used. However, I have found important to investigate the possible temperature difference $T_{diff}$ when the thermal conductivity is changed, accounting the fact that the material might be different in the real design. Thus thermal conductivity from $100 \frac{W}{mK}$ to $237 \frac{W}{mK}$ is considered, which are the realistic values for an aluminum alloy. The maximum temperature has a strong dependence on this material property as one can see in Figure \ref{fig:Tmax_TC_Absorber_Water}. Most likely, the material will be Al1050, which has a thermal conductivity of $222 \frac{W}{mK}$. The difference in maximum temperature between pure aluminum and Al1050 is $0.12~^\circ C$, which is small, and the calculations with pure aluminum are also a good approximation for Al1050 in steady-state situations.

\subsection{Concept B}

\subsubsection{Heat transfer coefficient}

\begin{figure}[!h]
\centering
\begin{minipage}{12cm}
\centering
\includegraphics[width=12cm]{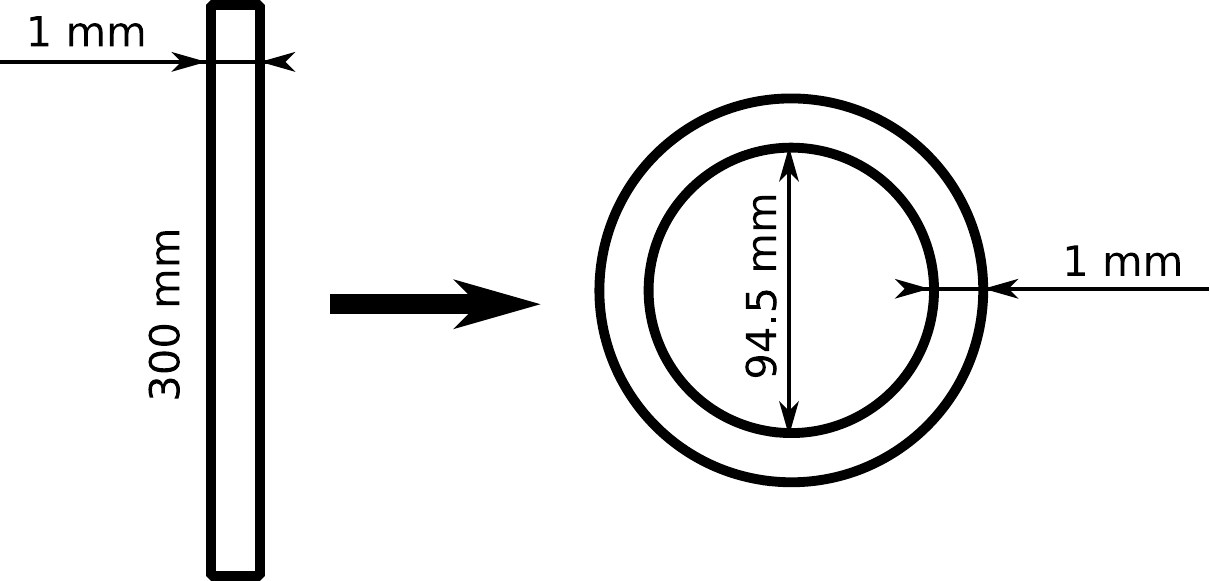}
\caption{Transformation between a rectangular cross-section gap and a narrow annular gap.}
\label{fig:Equ_Geom}
\end{minipage}
\end{figure}

The calculation of the heat transfer coefficient is not evident in this case. There is a wide and long, but very narrow gap between the layers. Considering it as a tube with an equivalent tube diameter does not work evidently. The heat transfer coefficient in narrow annular gaps was measured previously by Tachibana and Fukui \cite{HTC_article1}. 
Now, I am considering the equivalent dimension between a narrow annular gap and a gap with rectangular cross-section by wrapping the rectangular one around a cylinder (see Figure \ref{fig:Equ_Geom}). One can find the geometrical results in Table \ref{tab:equvalentgeom}. The equivalent tube diameter is necessary for the calculation of Reynolds (Re) number. It is calculated based on the dimensions of the annular gap, but one would obtain almost the same results using the rectangular cross-section directly.

\begin{table}[h!]
  \begin{center}
    \begin{tabular}{|l|r|r|r|} 
      \hline
      \textbf{Dimension} & \textbf{Notation} &  \textbf{Calculation} & \textbf{Value [mm]}\\
      \hline
      Hight of the air gap & $L_1$ & - & 200\\
      Width of the air gap & $L_2$ & - & 300\\
      Thickness of the air gap & $L_3$ & - & 1\\
      Equivalent outer diameter & $D_1$ & $\frac{L_2}{\pi} + L_3$ & 96,5\\
      Equivalent inner diameter & $D_2$ & $\frac{L_2}{\pi} - L_3$ & 94,5\\
      Equivalent tube diameter & $D_e$ & $\frac{(D_1^2 - D_2^2)\pi}{(D_1 + D_2)\pi}=D_1 - D_2=2L_3$ & 2\\
      \hline
    \end{tabular}
  \end{center}
  \caption{Geometry of the air gap between two layers and the equivalent annular air gap.}
  \label{tab:equvalentgeom}
\end{table}

The general strategy to determines the heat transfer coefficient requires an empirical (or semi-empirical) relation between the Nusselt (Nu) number (the dimensionless heat transfer coefficient) and some other dimensionless numbers such as the Reynolds (Re), Prandtl (Pr), Grashof (Gr), depending on the particular situation.

Here, I use the following relation to determine the Nusselt number \cite{HTC_article1}:
\begin{equation}
Nu=0.017\left(1+2.3\frac{D_e}{L}\right)\left(\frac{D_2}{D_1}\right)^{0.45} Re^{0.8} Pr^{1/3},
\end{equation}
where the Reynolds number and the Prandtl number are calculated using the equations \cite{hokozles}:

\begin{figure}[H]
        \centering
        \begin{minipage}{6cm}
        \begin{equation}
        Re=\frac{wD_e}{\nu}
        \end{equation}
        \end{minipage}
        \qquad
        \begin{minipage}{2cm}
        ~~~~~~~~~~~~~~and
        \end{minipage}
        \qquad
        \begin{minipage}{6cm}
        \begin{equation}
        Pr=\frac{\nu}{a}~.
        \end{equation}
        \end{minipage}
\end{figure}

\begin{figure}[h!]
\centering
\centering
\includegraphics[width=12cm]{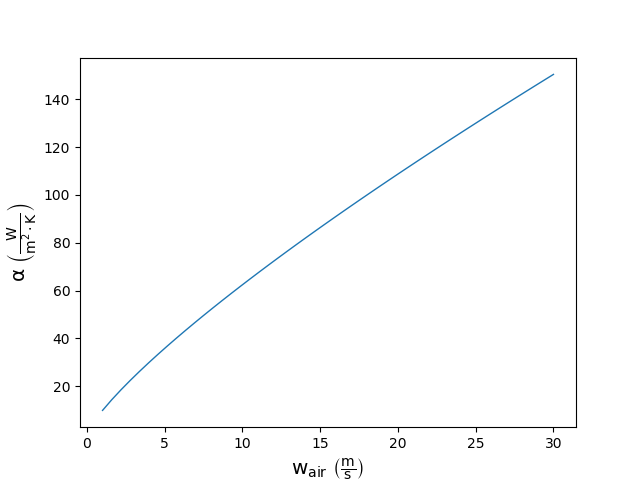}
\caption{The heat transfer coefficient as a function of airspeed.}
\label{fig:HTC_w}
\end{figure}

Using the Nu number, I can determine the heat transfer coefficient as
\begin{equation}
{\alpha}=\frac{Nu \lambda}{D_e}.
\end{equation}
For this particular situation, one can see the heat transfer coefficient as a function of airspeed in Figure \ref{fig:HTC_w}, and the constants are summarized in Table \ref{tab:constantsofair}.

\begin{table}[!h]
  \begin{center}
    \begin{tabular}{|l|r|r|r|} 
      \hline
      \textbf{Constant} & \textbf{Notation} &  \textbf{Value} & \textbf{Unit}\\
      \hline
      Air temperature & $T$ & $25$ & [$^\circ C$]\\
      Density & $\rho$ & $1.19$ & $\big[\frac{kg}{m^3}\big]$\\
      Kinematic viscosity & $\nu$ & $1.58 \cdot 10^{-5}$ & \big[$\frac{m^2}{s}\big]$\\
      Specific heat  & $c$ & $1013$ & $\big[\frac{J}{kg \cdot K}\big]$\\
      Thermal conductivity & $\lambda$ & $0.0261$ & $\big[\frac{W}{m \cdot K}\big]$\\
      Thermal diffusivity & $a$ & $2.17 \cdot 10^{-5}$ & $\big[\frac{m^2}{s}\big]$\\
      \hline
    \end{tabular}
  \end{center}
  \caption{Parameters of the air.}
  \label{tab:constantsofair}
\end{table}

\subsubsection{Temperature distribution}

\begin{figure}[h!]
\centering
\centering
\includegraphics[width=6cm]{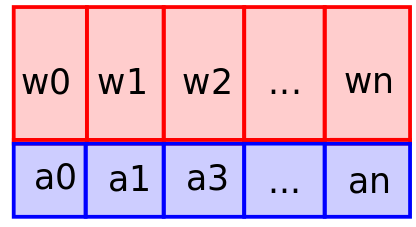}
\caption{The red cells denote the aluminum layer, the blue cells denote the air.}
\label{fig:Al_and_Air}
\end{figure}

In order to obtain the temperature distribution, I have applied the following simple numerical approximation and discretization. 
The solution is based on the internal energy balance, accounting the heat transfer among each cell. The aluminium plate and the air gap are splitted into $n$ parts, see Figure \ref{fig:Al_and_Air} for details. Every cell of the aluminum (`w-type' cell) are connected by heat conduction and connected by heat convection to the air (`a-type' cell). Heat conduction among the a-type cells is not considered. It is possible to determine the steady-state temperature distribution by iteration among the cells.

\begin{figure}[H]
\centering
\centering
\includegraphics[width=10cm]{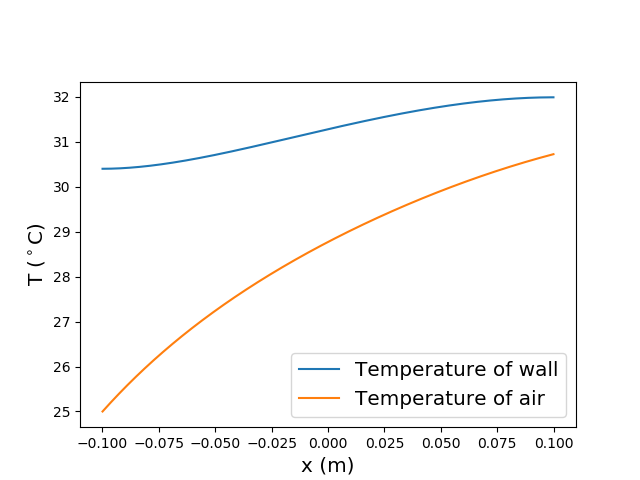}
\caption{The air and the layer temperature distribution in a streamline. The direction of x axis is visible in Figure \ref{fig:Air_Geom}. Average airspeed is set to 10 $\frac{m}{s}$, and the heat transfer coefficient is 62 $\frac{W}{m^2 \cdot K}$.}
\label{fig:T_air_w5}
\end{figure}

The results are presented in Figure \ref{fig:T_air_w5}. It makes apparent how the temperature of air increases between the entry and the exit points. More importantly, it influences the temperature distribution of a layer as well. It makes the heat transfer less intensive at the end, thus higher temperature difference can occur together with less homogeneous distribution. Consequently, the highest temperature can be found at the uttermost point from the entry of the air.

\subsubsection{Maximum temperature}

\begin{figure}[!h]
\centering
\centering
\includegraphics[width=11cm]{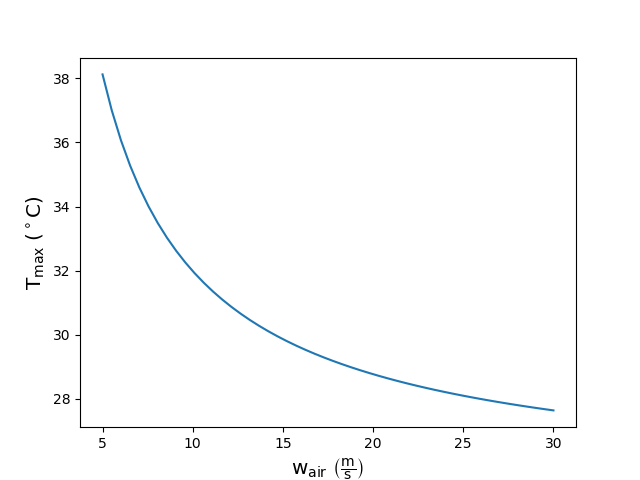}
\caption{Maximum temperature as a function of airspeed. The room temperature is 25 $^\circ C$.}
\label{fig:Tmax_airspeed}
\end{figure}

One can see the maximum temperature as a function of airspeed in the Figure \ref{fig:Tmax_airspeed}. The maximum temperature is lower than the allowed $40~^\circ C$ if the airspeed is higher than $5 \frac{m}{s}$. However, it is safer to ensure $15 \frac{m}{s}$ airspeed, but it is not worth to increase it. The optimum interval could be between $10$ to $15$ m/s. In the further calculations, $10$ m/s is used. Unfortunately, that speed could be too noisy and too high for clinical applications and may cause other problems as well.

\begin{figure}[!h]
\centering
\centering
\includegraphics[width=11cm]{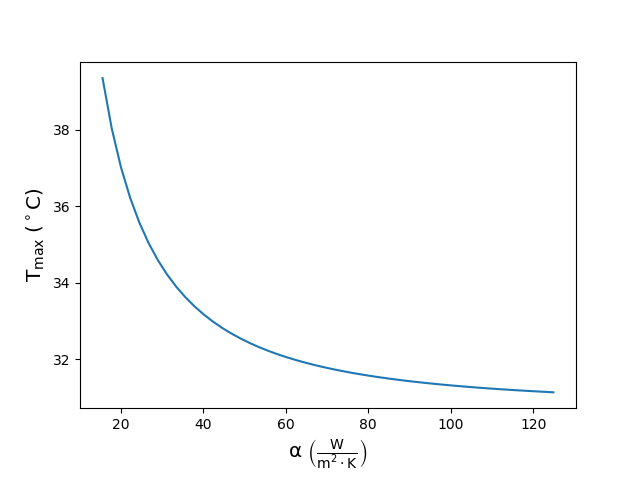}
\caption{The maximum temperature as a function of the heat transfer coefficient. The room temperature is set to 25 $^\circ C$.}
\label{fig:Tmax_HTC}
\end{figure}

One significant uncertainty is the determination of the heat transfer coefficient. The heat convection between the wall and airflow is a very complicated process because of the irregularity and uncertainty in the shape of the surface. Hence it is important to investigate the effect of changing the heat transfer coefficient. One can see the maximum temperature as a function of heat transfer coefficient in the Figure \ref{fig:Tmax_HTC}. The domain of interest starts at $40~\frac{W}{m^2 \cdot K}$ to $80~\frac{W}{m^2 \cdot K}$. Within this domain, the maximum temperature is safely lower than the allowed $40~^\circ C$.

\begin{figure}[H]
\centering
\centering
\includegraphics[width=10cm]{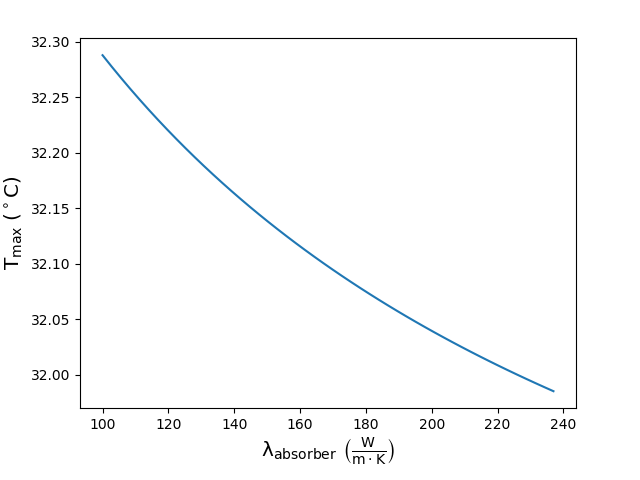}
\caption{The maximum temperature as a function of thermal conductivity.}
\label{fig:Tmax_TC}
\end{figure}

Previously, I mentioned the uncertainty due to the material change that should be investigated. Now, I handle the thermal conductivity of the absorber layer as a parameter, see Fig.~\ref{fig:Tmax_TC} for details. The domain of interest is $100 \frac{W}{mK}$ to $237 \frac{W}{mK}$ which covers all the possibilities for aluminium alloys. 
 The maximum temperature variation is found to be $0.3~^\circ C$ in this interval, hence the thermal conductivity of the absorber does not have a big impact on the maximum. One of the most likely aluminum alloy for the material of the detector is Al1050, which has the thermal conductivity of $222 \frac{W}{mK}$. The difference between the maximum temperature in case of pure aluminum and Al1050 is $0.02~^\circ C$, which is negligible.

\subsection{Comparison}

The maximum temperature difference in a layer is very important, since the cluster size and the noise rate of the ALPIDE are temperature dependent. I compared the maximum temperature differences for concepts A and B as a function of the airspeed, thermal conductivity and heat transfer coefficient. 

\begin{figure}[h!]
\centering
\centering
\includegraphics[width=10cm]{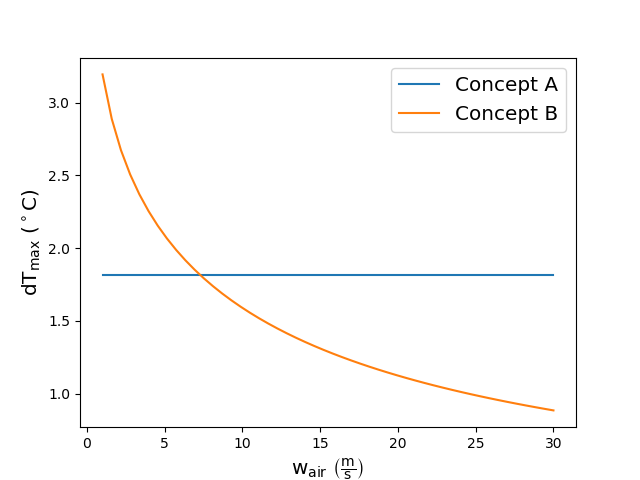}
\caption{The temperature difference in a  layer in the case of air and water cooling. The temperature difference in case of concept A is given as benchmark as it is independent of the airspeed.}
\label{fig:Tdiff_a_w}
\end{figure}

One can see the maximum temperature difference as a function of airspeed in case of air and water cooling in the Figure \ref{fig:Tdiff_a_w}. As one can see, there is no significant difference in this value. If the airspeed is higher than $8 \frac{m}{s}$ the temperature difference is slightly lower for concept B. Otherwise, the temperature difference is higher than for water cooling. The temperature difference in case of concept B, considering $10 \frac{m}{s}$ airseed, is $1.6~^\circ C$. The maximum temperature difference in case of concept A is given as a benchmark, and it is constantly $1.8~^\circ C$. In both cases, the temperature difference is lower than the allowed 5 $^\circ C$.

\begin{figure}[!h]
\centering
\centering
\includegraphics[width=10cm]{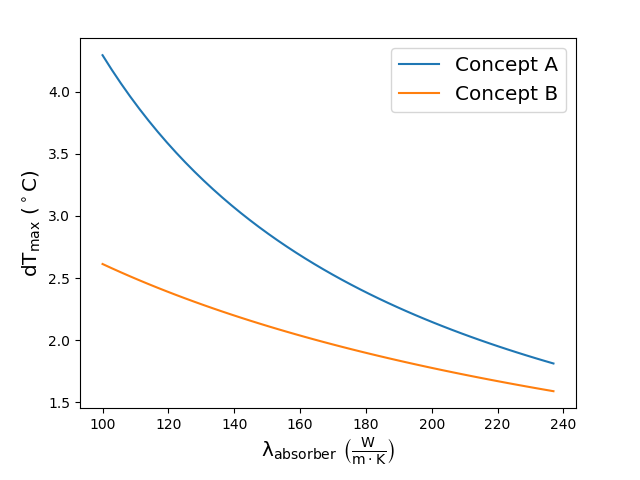}
\caption{Temperature difference in layer in case of air- and watercooling in function of the thermal conductivity of the layer. The average airspeed is set to 10 $\frac{m}{s}$ in case of concept B.}
\label{fig:Tdiff_a_w_TC}
\end{figure}

The maximum temperature difference depends on the thermal conductivity of the layer, as one can see in Figure \ref{fig:Tdiff_a_w_TC}, using $10 \frac{m}{s}$ airspeed. One can see that the temperature difference is increasing in both cases with decreasing the thermal conductivity. The concept A, using water cooling, seems to be more sensitive for this parameter. Considering again the Al1050 alloy, the temperature difference in the layer is $1.9~^\circ C$ for concept A and $1.7~^\circ C$ in case of concept B.

\begin{figure}[!h]
\centering
\centering
\includegraphics[width=10cm]{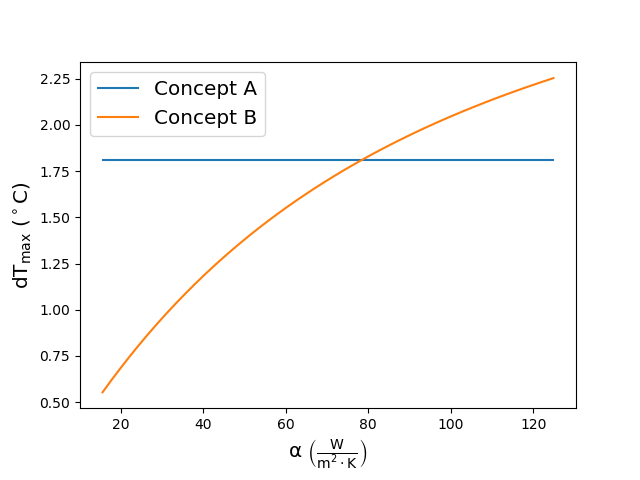}
\caption{The temperature difference in a layer in case of air and water cooling as a function of the heat transfer coefficient. The temperature difference in case of concept A is given as a benchmark.}
\label{fig:Tdiff_a_w_HTC}
\end{figure}

\begin{figure}[H]
\centering
\centering
\includegraphics[width=10cm]{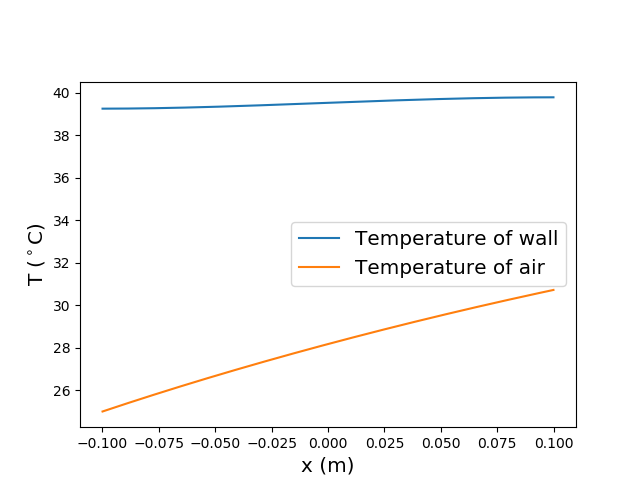}
\caption{The layer and the air temperature with unrealistically low (15 $\frac{W}{m^2 \cdot K}$) heat transfer coefficient, in stream direction. The direction of x axis is visible in Figure \ref{fig:Air_Geom}. The average air speed is set to 10 $\frac{m}{s}$.}
\label{fig:T_line_lowHTC}
\end{figure}

\begin{figure}[H]
\centering
\centering
\includegraphics[width=10cm]{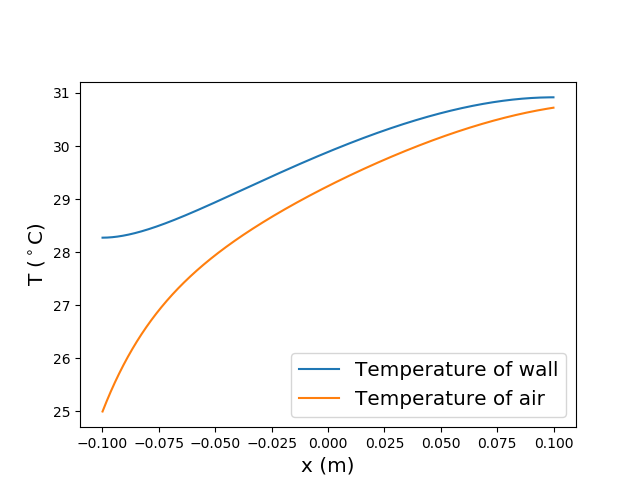}
\caption{The layer and the air temperature with unrealistically high (200 $\frac{W}{m^2 \cdot K}$) heat transfer coefficient, in stream direction. The direction of x axis is visible in Figure \ref{fig:Air_Geom}. The average air speed is set to 10 $\frac{m}{s}$.}
\label{fig:T_line_highHTC}
\end{figure}

Changes in the heat transfer coefficient between the layer and the air also has an effect in the maximum temperature difference. One can see the outcome in the Figure \ref{fig:Tdiff_a_w_HTC}. If the heat transfer coefficient is increasing the temperature difference is also increasing. The reason behind this effect is the changes in the temperature distribution in the streamline. In Figure \ref{fig:T_line_lowHTC}, one can see the temperature distribution using unrealistically low heat transfer coefficient, and in Figure \ref{fig:T_line_highHTC}, with unrealistically high  heat transfer coefficient. The realistic zone of the heat transfer coefficient is between $40~\frac{W}{m^2 \cdot K}$ and $80~\frac{W}{m^2 \cdot K}$.
In this domain, the maximum temperature difference starts from $1.2~^\circ C$ to $1.8~^\circ C$ in case of concept B. In case of concept A, the maximum temperature difference is not affected by the heat transfer coefficient, so it is only given as a benchmark, it is constantly $1.8~^\circ C$.

\newpage
\section{Cost estimation and environmental effects}

As a result of the previous calculations, I considered that both concept A and concept B meets with the requirements of the hadron-tracking calorimeter, so the decision between these concepts can be made based on other aspects. One of these aspects is the cost of the cooling system, so I compare the possible cost of concept A and concept B in this chapter. This comparison is a approximation of the costs, so it contains only the cost of the main elements of the cooling system. The prices in Table \ref{tab:costofA} and Table \ref{tab:costofB} are given as an example for the price of the parts.

\subsection{Concept A}

\begin{figure}[H]
\centering
\centering
\includegraphics[width=10cm]{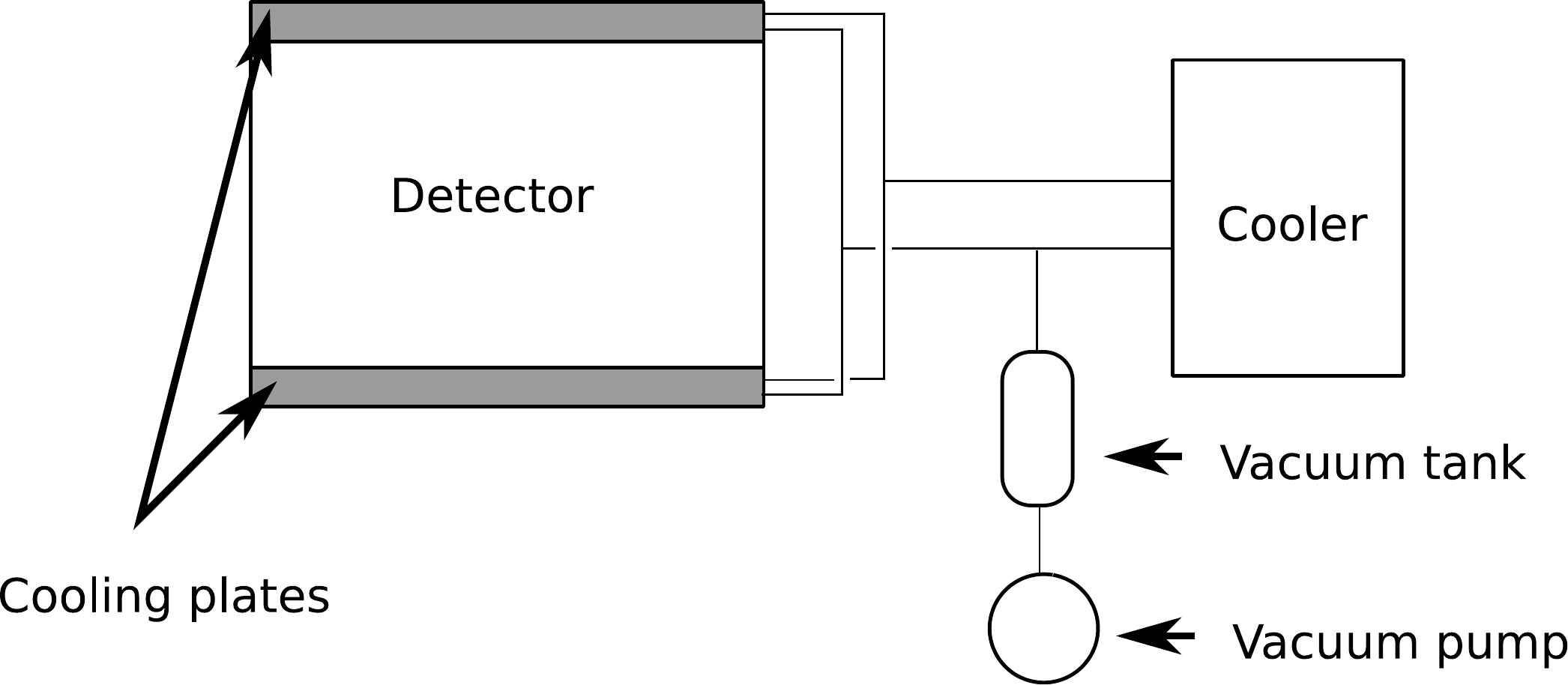}
\caption{The water cooling circuit of concept A.}
\label{fig:A_concept_gazd}
\end{figure}

In case of concept "A" I use a water cooling circle. To protect the hadron-tracking calorimeter from a possible leak, I need to use water cooling circle, which operates under room pressure. If the cooling system operates under room pressure, in case of a leak the air of the room move into the tube, not the water move out. To operate under room pressure I need to use a vacuum pump. In case of a leak the vacuum pump has to be able to remove all water from the cooling circle, so I need the insert a vacuum tank between the cooling circle and the vacuum pump. In normal working this tank mainly filled with air. If there is a leak in the cooling circle this tank will be filled with the water of the cooling circle. The cooling circle contains a cooler, to cool the water under the temperature of the room. The cooler contains a pump, which circulates the water in the cooling circle, so I did not calculate the price of the pump separately.  The cooling circle also contains cooling plates to cool the detector. I only find smaller cooling plates then the top or the bottom of the hadron-tracking calorimeter, so I calculated with two cooling plate in the top and two cooling plate in the bottom of the detector. The cooling circle also contains tubes and fittings. I just estimated price of them, because without an exact plan of the cooling system I could not find price for them. The elements of concept A is visible in Figure \ref{fig:A_concept_gazd}. One can see the estimated cost of this solution in Table \ref{tab:costofA}.

\begin{table}[!h]
  \begin{center}
    \begin{tabular}{|l|r|r|r|} 
      \hline
      \textbf{Component} & \textbf{Main operation data} &  \textbf{Number of components} & \textbf{Price}\\
      \hline
	Water cooler \cite{cooler} & $2400~W,~16.3~\frac{l}{min},~2.1~bar$ & 1 & 4090 \texteuro \\
	Vacuum pump \cite{vacuumpump} & 100 mbar & 1 & 640 \texteuro \\
	Cooling plate \cite{coolingplate} & 304.8 mm $\cdot$ 88.9 mm& 4 & 110 \texteuro \\
	Vacuum tank \cite{vacuumtank}  & 25 liter & 1 & 50 \texteuro \\
	Tubes and fittings & - & - & 300 \texteuro \\
	Overall & - & - & 5520 \texteuro \\
      \hline
    \end{tabular}
  \end{center}
  \caption{Cost of concept "A".}
  \label{tab:costofA}
\end{table}

\subsection{Concept B}

I need to calculate the volume flow through the detector and the pressure drop in the detector to find a corresponding fan to concept B.

Pressure drop in case of concept B, if the temperature of the room is $T_{room}=25~^\circ C$. The equivalent diameter is:
\begin{equation}
d_{eq}=4 \cdot \frac{L_2 \cdot L_3}{2 \cdot L_2 + 2 \cdot L_3}=1.99~mm,
\end{equation}
 where $L_2=300~mm$ and $L_3=1~mm$. The airspeed is $w_{air}=10~\frac{m}{s}$, and the kinematic viscosity is $\nu=1.58 \cdot 10^{-5}$ $\frac{m^2}{s}$. The Reynolds-number of the airflow is:
\begin{equation}
Re=\frac{w_{air} \cdot d_{eq}}{\nu}=1250,
\end{equation}
which means the airflow is laminar between the detector layers. The pressure drop can be calculated with this equation
\begin{equation}
\Delta p=\frac{3 \cdot L_1 \nu \cdot \rho w_{air}}{\left(\frac{L_3}{2}\right)^2}=438~Pa
\end{equation}
based on Newton's law of viscosity, where $\rho=1.19~\frac{kg}{m^3}$ and $L_1=200~mm$. The volume flow between the layers is
\begin{equation}
Q=n_{layers} \cdot L_3 \cdot L_2 \cdot w_{air}=0.105~\frac{m^3}{s}=378~\frac{m^3}{h},
\end{equation}
where $n_{layers}=35$.

\begin{figure}[H]
\centering
\centering
\includegraphics[width=3.5cm]{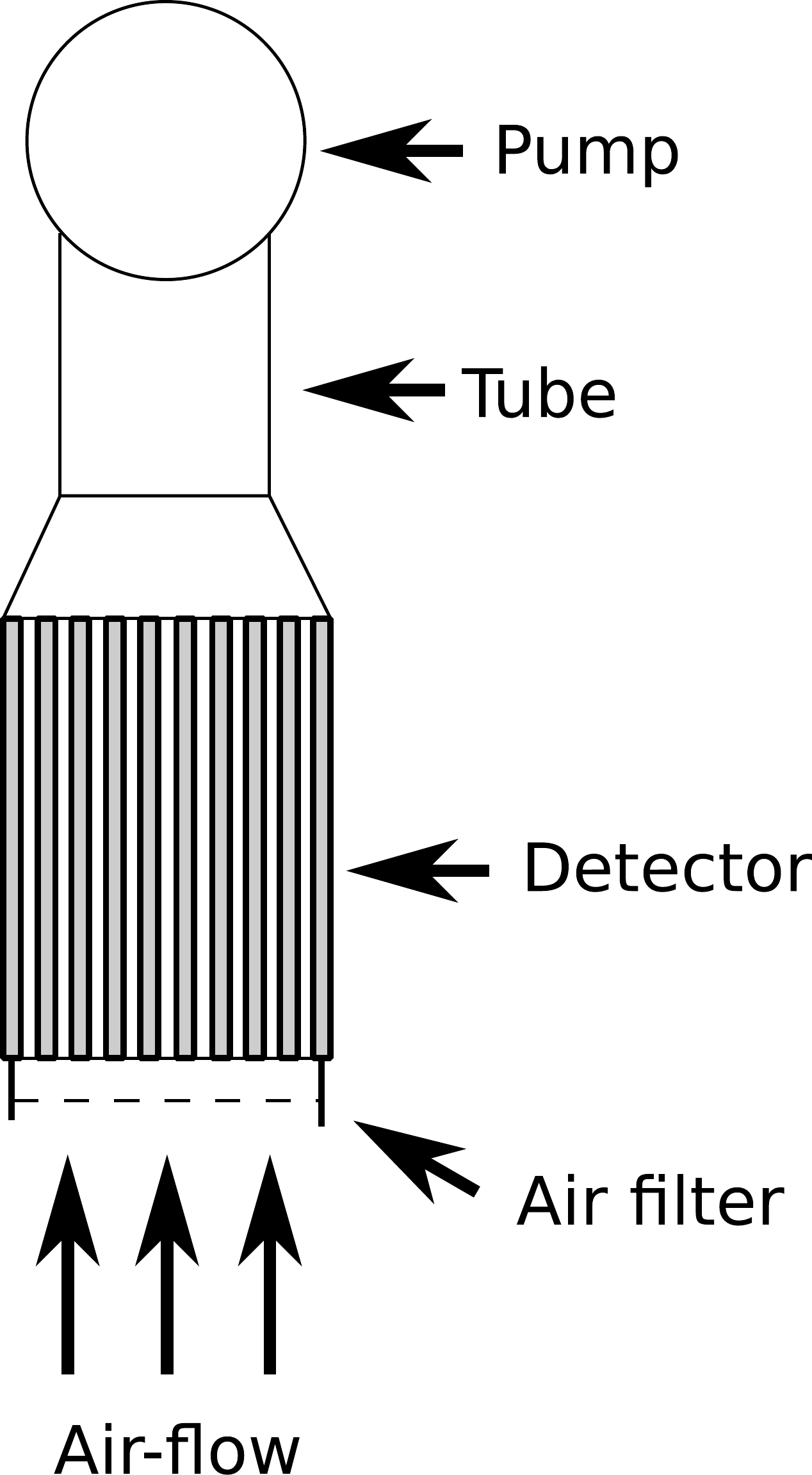}
\caption{The elements of B cooling system concept.}
\label{fig:B_concept_gazd}
\end{figure}

I need to have a fun in the cooling system to generate airflow in the detector. The fan is as big as the hadron-tracking calorimeter and it is possible the fan generate vibrations, so it has to operate separately from the detector connected with a tube. I also need an air filter before the detector, to protect the hadron-tracking calorimeter from the dust in the air. The elements of the cooling system is visible in Figure \ref{fig:B_concept_gazd}.

\begin{table}[!h]
  \begin{center}
    \begin{tabular}{|l|r|r|r|} 
      \hline
      \textbf{Component} & \textbf{Main operation data} &  \textbf{Number of components} & \textbf{Price}\\
      \hline
	Fan \cite{fan} & 950 Pa, $400~\frac{m^3}{h}, 485 W$ & 1 & 890 \texteuro \\
	Tube \cite{tube} & l=10 m, d=250 mm & 1 & 50 \texteuro \\
	Fittings and filter & - & - & 300 \texteuro \\
	Overall & - & - & 1240 \texteuro \\
      \hline
    \end{tabular}
  \end{center}
  \caption{Cost of concept "B".}
  \label{tab:costofB}
\end{table}

\subsection{Comparison}

As a result of the cost estimation one can conclude that the concept B is cheaper than concept A.  As environmental effect one can consider that concept B use less energy to operate than concept A, because the water cooler needs more energy to cool the water under room temperature, than the fan needs to transfer air through the detector.

It is possible to use a concept like A, but use thermoelectric cooler and heat sinks with heat pipe and funs to cool the top and the bottom of the detector. This solution results same temperature distribution than concept A, but it can be cheaper, smaller than concept A. In case of this concept it is also possible to cool the detector under room temperature and the maximum temperature also adjustable. This concept can be more advantageous than concept A in case of beam test, but concept A is more suitable for clinical use, because in this case the generated heat is transported out of the treatment room.

\newpage
\section{Finite element simulation of concept A}

My next work in the research team was the more accurate calculation of the temperature distribution in case of concept A. I decided to check the effect of contact thermal resistances and inhomogeneous load on the temperature distribution. I took into account the heat generation of the ALPIDEs as a heat flux in the contact surfaces of them. I calculated with a two dimensional model (Figure \ref{fig:Tresistance_layer_structure}) when I calculate the effect of the contact resistances, and I calculate with a three dimensional model (Figure \ref{fig:Stacks}) when I calculated the effect of the inhomogeneous load. I used $\lambda_{Al1050}=222~\frac{W}{m \cdot K}$ as the thermal conductivity of the absorber layers in both calculation.

\subsection{The effect of contact thermal resistances}

\begin{figure}[H]
\centering
\centering
\includegraphics[width=10cm]{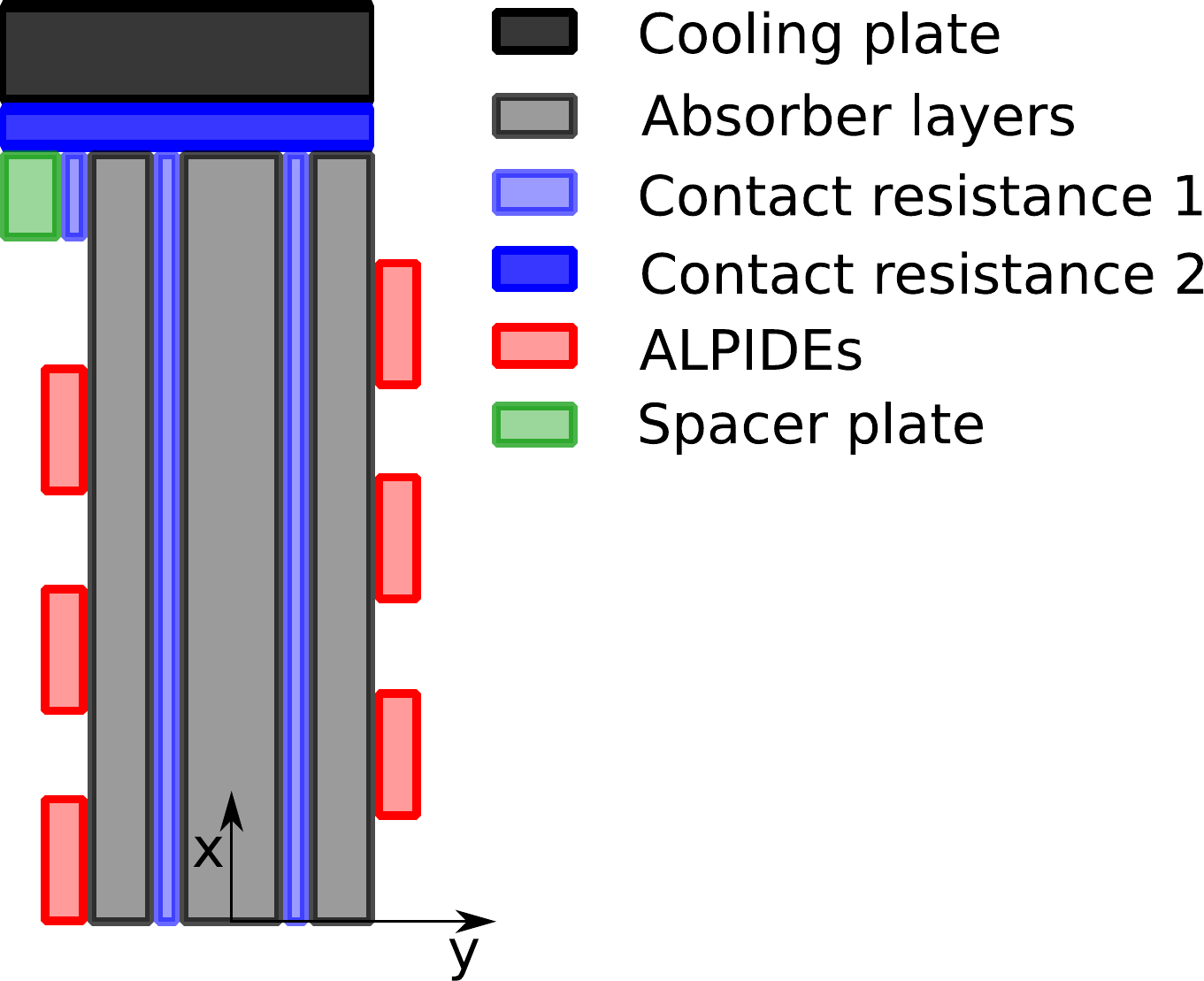}
\caption{The model of the layer with thermal resistances.}
\label{fig:Tresistance_layer_structure}
\end{figure}

Firstly I calculated with 4 mm thick absorber layer. In reality the layer will be build from 2 mm thick absorber layer in the middle, and two 1 mm thick layer in both side. The ALPIDEs will be glued into the 1 mm thick plates, as one can see in \ref{fig:Tresistance_layer_structure}. There will be contact resistances between this plates and between the cooling plate and this plates. I calculate the contact resistance between the plates as the resistance of the average air gap between the layers, based on \cite{contactRes}. I calculate the average air gap between two plate with $\delta_1=\frac{2 \cdot R_z}{2}=R_z$ equation. The aluminum plate is cold rolled aluminum, so the mean roughness depth is $R_z=12.5~\mu m$. The average distance is $\delta_1=12.5~\mu m$. Between the cooling plate and the edge of the absolber plates I calculated with $\delta_2=200~\mu m$, because of the uncertainty of the positioning of the plates. It is necessary to use thermal conductive grease between the cooling plate and the edge of the absolber plates. I considered $\lambda_{air}=0.0261~\frac{W}{m \cdot K}$ as the thermal conductivity of air and $\lambda_{conductive}=8.5~\frac{W}{m \cdot K}$ \cite{thermalpaste} as the thermal conductivity of a typical thermal conductive paste. The thermal resistance between two layer is $R_1=\frac{\delta_1}{\lambda_{air}}=4.79 \cdot 10^{-4}~\frac{m^2 \cdot K}{W}$ and between the cooling plate and the edge of the absorber layers is $R_2=\frac{\delta_2}{\lambda_{conductive}}=2.35 \cdot 10^{-5}~\frac{m^2 \cdot K}{W}$.

\begin{figure}[H]
\centering
\centering
\includegraphics[width=10cm]{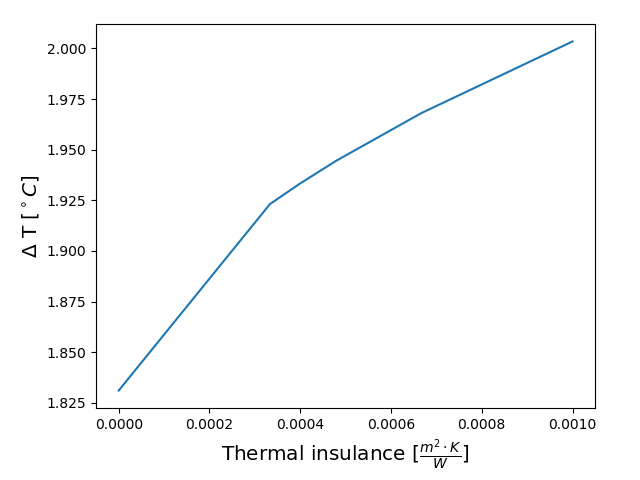}
\caption{The effect of the thermal resistance between layers in the maximum temperature difference.}
\label{fig:Tresistance_results1}
\end{figure}

\begin{figure}[H]
\centering
\centering
\includegraphics[width=10cm]{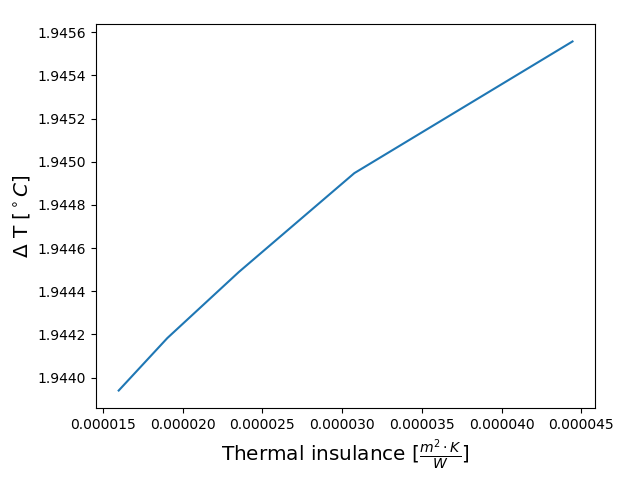}
\caption{The effect of the thermal resistance between the cooling plate and absorber the layers in the maximum temperature difference.}
\label{fig:Tresistance_results2}
\end{figure}

One can see the model of the layer and the contact resistances in Figure \ref{fig:Tresistance_layer_structure} and the result of the two dimensional simulation in Figure \ref{fig:Tresistance_results1} and Figure \ref{fig:Tresistance_results2}. I calculated the maximum temperature difference in that part of the surface of the absorber layer, where ALPIDEs are glued.

I have checked the mesh dependence of the solution in case when  $R_1$ is $4.79 \cdot 10^{-4}~\frac{m^2 \cdot K}{W}$ and $R_2$ is $2.35 \cdot 10^{-5}~\frac{m^2 \cdot K}{W}$. I checked the maximum temperature difference ($\Delta T_{max}$) in y=2 mm line, which is visible in Table \ref{tab:Tresistance_mesh}. I found that the mesh with 0.5 mm edge length seems to be the most accurate. The maximum temperature difference between 1 mm edge length and 0.5 mm edge length is small compared with the measured temperature differences, so the solution is mesh independent.

\begin{table}[H]
  \begin{center}
    \begin{tabular}{|r|r|r|r|} 
      \hline
      $L_{max}$ in x direction (mm) & $L_{max}$ in y direction (mm) & Number of elements & $\Delta T_{max}~(^\circ C)$\\
      \hline
	1 & 1 & 800 & - \\
	0.5 & 0.5 & 3200 & 0.017\\
	2 & 0.5 & 824 & - \\
	1 & 0.25 & 3216 & 0.104\\
      \hline
    \end{tabular}
  \end{center}
  \caption{Outcome of the mesh dependence test, where $\Delta T_{max}$ is the maximum difference of temperature compared with the previous mesh. $L_{max}$ is the maximum edge length in the given direction.}
  \label{tab:Tresistance_mesh}
\end{table}

I considered based on Figure \ref{fig:Tresistance_results1} and Figure \ref{fig:Tresistance_results2} that the contact thermal resistance between the layers has minimal effect on the temperature difference, and the contact thermal resistance between the cooling plate and the layers has only marginal effect on the temperature distribution.

\subsection{The effect of inhomogeneous load}

I obtained the temperature distribution in the case, if only the middle of the detector is used. I expect the maximum temperature difference will be smaller than in case of full load, but it is interesting how much it will be, because the partial load will be a typical usage, if we have to take image from a smaller part of the human body then the size of the full detector area.

I used $P_{standby}=60~mW$ heat generation in every ALPIDE detector, which was inactive in the measurement, which is the power consumption of waiting for hits. I used $P_{active}=185~mW$ heat generation in every active ALPIDE, which is the heat generation working with maximum readout capacity.

\begin{figure}[H]
\centering
\centering
\includegraphics[width=10cm]{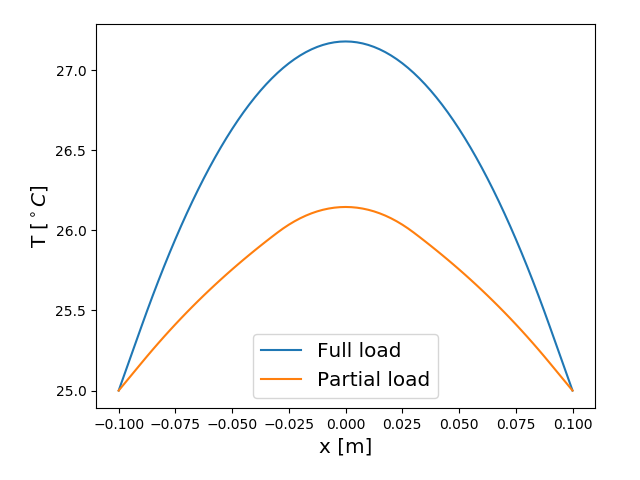}
\caption{The temperature distribution in case of full and partial load in path z=0 mm and y=0.002 mm.}
\label{fig:T_x_partial_load1}
\end{figure}

I obtained the temperature distribution in an x direction path, defined by z=0 mm and y=0.002 mm. The temperature distribution in this path is visible compared with the temperature distribution of full load in Figure \ref{fig:T_x_partial_load1}.

\begin{figure}[H]
\centering
\centering
\includegraphics[width=10cm]{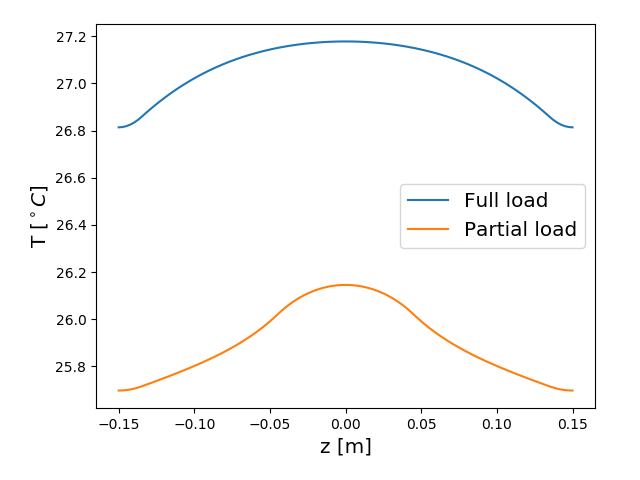}
\caption{The temperature distribution in case of full and partial load in path x=0 mm and y=0.002 mm.}
\label{fig:T_x_partial_load2}
\end{figure}

One can see the temperature in a path, defined by x=0 mm and y=0.002 mm, in Figure \ref{fig:T_x_partial_load2}.

The result of the mesh dependence test is visible in Table \ref{tab:T_inhomogen_mesh}.

\begin{table}[H]
  \begin{center}
    \begin{tabular}{|r|r|r|r|} 
      \hline
      Maximum edge size (mm) & Number of elements & $\Delta T_{max}~(^\circ C)$ full load & $\Delta T_{max}~(^\circ C)$ full load \\
      \hline
	7.5 & 1234 & - & - \\
	5 & 2400 & 0.171 & 0.077 \\
	2.5 & 19200 & 0.086 & 0.038 \\
      \hline
    \end{tabular}
  \end{center}
  \caption{Outcome of the mesh dependence test in case of full and partial load, where the $\Delta T_{max}$ is the maximum temperature difference compared with the previous mesh.}
  \label{tab:T_inhomogen_mesh}
\end{table}

The temperature difference was $2.1~^\circ C$ in case of full load and $1.1~^\circ C$ in case of partial load.

\newpage
\section{Calculation of transient behavior}

In this subsection I focus on how much time does the heat up takes, because it effects both the clinical use, both the measurements during the development of the detector. I used the same geometry than I used in Chapter 3.1, so I used a one dimensional model. I calculated with the thermal conductivity of Al1050, which is $222~\frac{W}{m \cdot K}$. The heat transfer is described by the following equation:
\begin{equation}
\label{mainEq}
\frac{\partial T}{\partial t}=\alpha \cdot \frac{\partial^2 T}{\partial x^2} + \frac{\dot q_v}{\rho \cdot c},
\end{equation}
where $\alpha=\frac{\lambda}{\rho \cdot c}$.

Initial condition:
\begin{equation}
\label{eqn:init1}
T(x,0)=T_0=25~^\circ C.
\end{equation}
Boundary conditions:
\begin{figure}[H]
        \centering
        \begin{minipage}{6cm}
        \begin{equation}
	T(0,t)=T_1=25~^\circ C
        \end{equation}
        \end{minipage}
        \qquad
        \begin{minipage}{2cm}
        ~~~~~~~~~~~~~~and
        \end{minipage}
        \qquad
        \begin{minipage}{6cm}
        \begin{equation}
	T(0,t)=T_2=25~^\circ C.
        \end{equation}
        \end{minipage}
\end{figure}

The temperature can be calculated as the sum of steady state and the transient temperature:
\begin{equation}
T(x,t)=T_s(x) + T_h(x,t).
\end{equation}
The steady state temperature is \cite{hokozles}:
\begin{equation}
T_s(x)=-\frac{\dot q_v}{\lambda} \cdot \frac{x^2}{2} + C_1 \cdot x + C_2,
\end{equation}
where $C_2=T_1$ and $C_1=\frac{T_2 - T_1}{L} + \frac{\dot q_v}{\lambda} \cdot \frac{L}{2}$. The transient temperature is \cite{hokozles}:
\begin{equation}
T_h(x,t)=\sum_{n=1}^{n_{max}} D_n \cdot e^{-\Lambda_n \cdot t} \cdot sin \left(\frac{n \cdot \pi}{L} \cdot x \right),
\end{equation}
where $\Lambda_n=\alpha \cdot \left( \frac{n \cdot \pi}{L} \right)^2$ and $D_n=\frac{\int_{0}^{L} (T(x,0)-T_s(x)) \cdot sin \left( \frac{n \cdot \pi}{L} \cdot x \right) dx}{\int_{0}^{L} sin^2 \left( \frac{n \cdot \pi}{L} \cdot x \right) dx}$.

\begin{figure}[H]
\centering
\centering
\includegraphics[width=10cm]{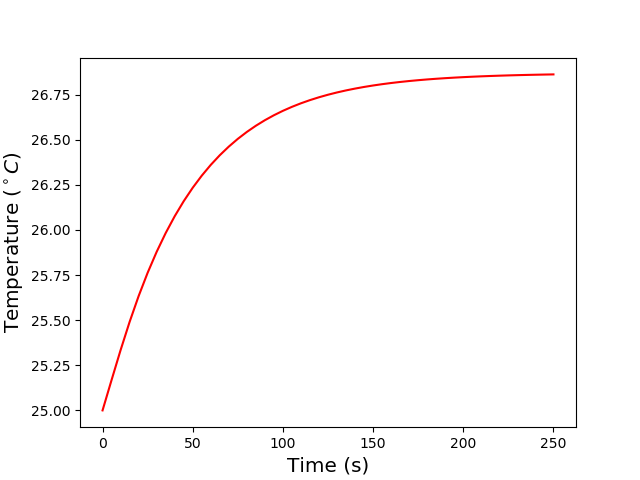}
\caption{The temperature of the middle point of the detector in function of time.}
\label{fig:Tmiddle_t}
\end{figure}

One can see the temperature of the middle of the detector in function of time in Figure \ref{fig:Tmiddle_t}, and the parameters, which was used to the calculation, in Table \ref{tab:consHeatup}.

\begin{table}[H]
  \begin{center}
    \begin{tabular}{|r|r|r|} 
      \hline
      Parameter & Value\\
      \hline
	Initial temperature of the layer ($T_1$) & $25~^\circ C$\\
	Temperature of the edge of the layer ($T_1$ and $T_2$) & $25~^\circ C$\\
	Thermal conductivity of the layer ($\lambda$) & $222~\frac{W}{m \cdot K}$ \\
	Mass density of the layer ($\rho$) & $2710~\frac{kg}{m^3}$ \\
	Specific heat of the layer (c) & $434~\frac{J}{kg \cdot K}$ \\
	Equivalent volumetric heat generation ($q_v$) & $8.3 \cdot 10^4~\frac{W}{m^3}$ \\
	Maximal item of Fourier series ($n_{max}$) & 50 \\
      \hline
    \end{tabular}
  \end{center}
  \caption{Parameters used in the heat up calculation.}
  \label{tab:consHeatup}
\end{table}

It takes 134 s, while the temperature of the middle point of the layer reach $\pm 0.1~^\circ C$ zone around the steady state temperature of the middle point.

\newpage
\section{Looking Ahead}

In this work I analysed the temperature distribution of a layer of the hadron-tracking calorimeter. The detector is going to contain two layer before the calorimeter, which measure the path of the particles. This layers are going to contains a carbon composite layer in the middle instead of an aluminum absorber layer. This carbon composite layer is going to work as a support structure, and it is thinner than aluminum absorber layer. The thermal conductivity of carbon composite is anisotropic, it is reasonable in longitudinal direction, but low in perpendicular direction. It is essential to analyze the temperature distribution in this layers, and design new cooling concepts for these layers, if it is necessary.

Another important engineering task in the development of  proton computer tomograph is the comparison of the temperature distribution between layers. This task requires more information about the power consumption of the tracking detectors and the number of particles, which goes through them. It is possible to reach this information with test beam measurements, so one of the next engineering task should be the measurement of the electric power consumption in case of different particle fluxes.

\newpage
\section{Summary}

Cancer started to become the leading cause of death in the developed world, as it is responsible for 25\% of all death in Hungary. Mainly, there are three ways of treatment: surgery, radiotherapy, and chemotherapy. Hadron therapy is a novel treatment in radiotherapy. It is beneficial because it generates less radiation in healthy tissues than X-ray radiotherapy, thus it leads to fewer side effects, and it allows a higher daily dose, which increases the effectiveness of treatment.

Hadron therapy requires an accurate three-dimensional map about the electron density of the body. The most accurate solution to obtain the map uses hadrons, however, it requires the development of a proton computer tomograph (pCT). Our group, the Bergen pCT Collaboration, aims to develop a pCT that can meet the requirements of the clinical use. The detector of the pCT is built of silicon pixel tracking detectors and aluminum absorber layers.

In the collaboration, my role was to investigate the temperature distribution of the detector, which is essential because of the accuracy of the energy measurement based on the temperature of the detector. It is indispensable in the development of the pCT detector.

I applied the lumped capacitance method to estimate the thermal behaviour of the detector, containing 35 absorber layers. This simple analysis shows the necessity for cooling. Hence, two cooling concepts are elaborated in this work. In concept A, water cools the edges of the aluminum absorber layers. In concept B, the air is circulated between the detector layers. My goal was to quantify the differences in the steady-state temperature distribution and compare the concepts to each other.

Firstly I calculated with analytic methods the steady state temperature distribution in case of both concepts. Both of the them satisfies the first requirement, which is to keep the maximum temperature lower than 40  $^\circ C$. The allowed temperature difference is 5 $^\circ C$. Both of the concepts meets this requirement also, but concept B has a bit better performance as the temperature difference is 1.6 $^\circ C$ in this case, compared with 1.8 $^\circ C$ in case of concept A.

Secondly I estimated the cost of both concept. I considered concept A is more expensive than concept B. It is possible to use a third solution, which cools the top and the bottom of the hadron-tracking calorimeter, like concept A, but contains thermoelectric coolers, heat pipes, heat sinks and funs instead of a water cooling circuit. This concept can be smaller and cheaper than concept A, but it has the advantageous properties of concept A.

Thirdly I used finite element simulations to determinate the effect of the contact thermal resistances and partial load in the temperature distribution, in case of concept A. I obtained that a reasonable contact thermal resistance has small effect on the temperature distribution, as it cause maximum $0.2~^\circ C$ increase in the temperature difference. In case of partial load the temperature difference is decrease, as it was expected. I calculated the time, which is required to reach steady state temperature distribution. It takes 134 s, which is suitable for clinical use and test beam measurements.

\newpage
\section{Összefoglaló}

Napjainkban a rák az egyik vezető halálokká kezd vállni a fejledt világban, Magyarországon a teljes elhalálozás 25\%-áért felelős. A ráknak három féle fő gyógy\-mód\-ja van: az első a sebészet, a második a sugárterápia és a harmadik a kemoterápia. A hadronterápia egy új sugárterápiás kezelés. Előnyös, mert kevesebb ionizációt okoz az egészséges szövetekben mint a hagyományos, röntgen sugárzással végzet sugárterápia. Ennek köszönhetően kevesebb mellékhatással jár, és nagyobb napi dózist tesz lehetővé, ezzel növelve a kezelés hatásosságát.

A hadronterápiához szükséges egy pontos térbeli képet készíteni a páciens testének elektronsűrűségéről. A legpontosabb megoldás, ha ezt a képet hadronok használatával készítjük el, de ez egy proton computer tomográf (pCT) kifejlesztését igényli. A Bergen pCT együttműködés célja, hogy kifejlesszen egy ilyen eszközt, ami a klinikai felhasználási szempontoknak eleget tesz. A pCT detektora szilíciumpixel-nyomkövető detektorokból és alumínum abszorber rétegekből épül fel.

Az együttműködésben betöltött szerepem a detektor hőmérséklet-eloszlásának meg\-ha\-tá\-ro\-zá\-sá\-ra irányult. Ennek vizsgálata elengedhetetlen a pontos mérések elvégzéséhez. Először, egy egy\-sze\-rű, koncentrált paraméterű modellel becsültem a 35 rétegből álló detektor termikus viselkedését. Ebből egyértelműen látni, hogy a hűtés elengedhetetlen.
Két féle hűtési koncepciót dolgoztam ki. Az ``A'' koncepció esetén vízzel hűtjöm az abszorber rétegek széleit, míg a ``B'' koncepció esetén levegőt keringtetek a rétegek között. Célom a kialakuló, állandósult állapotú hőmérséklet-eloszlások meghatározása és a két koncepció összehasonlítása.

Először analitikusan meghatároztam az állandósult hőmérsékleteloszlást mindkét koncepció esetében. A megengedett maximum hőmérséklet a detekteorban $40~^\circ C$, amit mindkét koncepció teljesít. A megengedett hőmérséklet-különbség $5~^\circ C$, amit úgyszintén mindkét koncepció teljesít, de a B koncepció ilyen szempontból előnyösebb, mivel ebben az esetben a hőmérséklet-különbség csak $1.6~^\circ C$, amíg az A koncepció esetében $1.8~^\circ C$.

Másodszor megbecsültem a két koncepció lehetséges költségeit. Eredményül azt kaptam, hogy az A koncepció lényegesen drágább. Lehetséges egy harmadik köncepció alkalmazás is, amely esetében ugyanúgy a detektor alsó és felső felületét hűtjúk, mint az A koncepció esetében, azonban vízhűtés helyett Peltier-elemek, hőcsövek, bordák és ventillátorok segítségével valósítjuk meg a hűtést. Ez a koncepció kisebb és olcsóbb megoldást jelenthet az A koncepciónál, azonban rendelkezik annak az előnyös tulajdonságaival.

Harmadszor végeselemes-szimulációk segítségével vizsgáltam a kontakt-hőellenállások és az részleges terhelés hatását a hőmérséklet eloszlásra az A koncepció esetében. Eredményül azt kaptam, hogy a reális kontakt-hőellenállások maximum $0.2~^\circ C$ hőmérsékletkülönbség-növekedést okoz. Részleges terhelés esetén a maximális hőmérséklet-különbség  kisebb, mint teljes terhelés esetén, ahogyan ezt előre vártuk. Analitikus számítással meghatároztam az állandósult hőmérsékleteloszlás kialakulásához szükséges időt. Ez 134 másodpercnek adódott, ami megfelelő, mind az orvosi felhasználás, mind a tesztmérések számára.

\newpage
\addcontentsline{toc}{section}{Acknowledgements}
\section*{Acknowledgements}

I would like to thank for the work of my supervisor, Róbert Kovács, and my consultant Mónika Varga-Kőfaragó. Both of them helped me a lot. First, I would like to thank to Mónika, who found this topic for me, and helped a lot in the initial difficulties. I studied a lot from her about data analysis, programming, and how to behave as a scientist while we worked together. I would like to thank Róbert for his help that I got from him during the creation of this thesis. I studied a lot from him about thermodynamics, and about how to work as an engineer. I hope we will continue the work together, and I can learn a lot more from him.

I would like to thank Gergely Gábor Barnaföldi, who helped me a lot with his advice and encouragement. I started to learn from him how to write not just grammatically correct, but readable sentences, that will be useful in my entire career to become a scientist.
I would like to thank Dieter Rörich, who gave me the possibility to spend three months in Bergen and study a lot from him and his colleagues. I also would like to thank Shruti Vineet Mehendale, who helped me a lot in my work during the last half-year. I owe a lot of gratitude for the Wigner \-ALICE Group and the Bergen pCT Group for the possibility of joining to their project. Finally, this work would not be possible without the support of NKFIH/OTKA K 120660, Hungarian Scientific Research Fund – OTKA.

\end{document}